\documentclass[cernpreprint, UKenglish, texmf, orcidlogo]{atlasdoc}
\usepackage{atlaspackage}
\usepackage{atlasbiblatex}
 
\usepackage{atlasphysics}
 
\graphicspath{{logos/}{figures/}}
 
\usepackage{ANA-TOPQ-2023-28-PAPER-defs}

\AtlasTitle{Precise test of lepton flavour universality in $\boldmath W$-boson decays into muons and electrons in $\boldmath pp$ collisions at $\boldmath\sqrt{s}=13\,\TeV$ with the ATLAS detector}

\AtlasAbstract{
The ratio of branching ratios of the $W$ boson to muons and electrons,
$\rwme=\brwm$/\linebreak[0]$\brwe$, has been measured using $\intlumi\,\ifb$ of
$pp$ collision data at \sxyt\ collected with the ATLAS detector at the LHC,
probing the universality of lepton couplings. The ratio is obtained from
measurements of the \ttbar\
production cross-section in the $ee$, $e\mu$ and $\mu\mu$ dilepton final
states. To reduce systematic uncertainties, it is normalised by the square root
of the corresponding ratio \rzme\ for the $Z$ boson
measured in inclusive \zee\ and \zmm\ events.
By using the precise value of \rzme\
determined from $e^+e^-$ colliders, the ratio \rwme\ is determined to be
\begin{equation*}
\rwme = \rwmeval \pm \rwmestat\,\mathrm{(stat)}\,\pm \rwmesyst\,\mathrm{(syst)}\,\pm \rwmeext\,\mathrm{(ext)} \ .
\end{equation*}
The three uncertainties correspond to data statistics, experimental
systematics and the external measurement of \rzme, giving a total uncertainty
of \rwmetot, and confirming the Standard Model assumption of
lepton flavour universality in $W$-boson decays at the 0.5\% level.
}

\AtlasRefCode{TOPQ-2023-28}
 
\PreprintIdNumber{CERN-EP-2024-063}

\AtlasJournalRef{Eur.\ Phys.\ J. C 84 (2024) 993}
\AtlasDOI{10.1140/epjc/s10052-024-13070-4}

\hypersetup{pdftitle={ATLAS document},pdfauthor={The ATLAS Collaboration}}
 
\begin{document}
 
\maketitle
 
\tableofcontents
 
\section{Introduction}\label{s:intro}
 
The assumption of lepton flavour universality, i.e.\ that the couplings of
the charged leptons $e$, $\mu$ and $\tau$ to the electroweak gauge bosons are
independent of the lepton masses, is a key axiom of the
Standard Model of particle physics. This assumption has been tested over
a wide range of momentum transfers by studying ratios of partial decay
widths (or equivalently, ratios of branching ratios) of various particles
to electrons, muons and taus. After correction for mass, phase space and
radiative effects, these ratios of decays into leptons of generations
$i$ and $j$ are proportional to $g_i^2/g_j^2$, where $g_i$ is the coupling
of lepton $i$ ($=e$, $\mu$, $\tau$). The equality of these couplings
has been tested to the 0.1--0.2\% level in decays of $\tau$ leptons,
$\pi$ and $K$ mesons (see for example Ref.~\cite{pichtau}).
More recently, hints of departures from lepton flavour universality
at the level of a few standard deviations were seen in the so-called
flavour anomalies in $b$-hadron decays,
e.g.\ in the processes $B\rightarrow D^{(*)}\tau\nu$ vs.\
$B\rightarrow D^{(*)}\ell\nu$ (with $\ell=e$ or
$\mu$)~\cite{babarbdtau,bellebdhad,bellebdpol,belle2bdhad,lhcbbd1,lhcbbd2},
and in the loop-induced
process $b\rightarrow s\ell\ell$. However, the latest measurement of
$b\rightarrow s\ell\ell$ in $B\rightarrow K^{(*)}\mu^+\mu^-$ vs.\
$B\rightarrow K^{(*)}e^+e^-$ decays from the LHCb collaboration
is in agreement with lepton flavour universality~\cite{lhcbbsll},
and definitive conclusions have yet to be established.
 
At high momentum transfer, the branching ratios for the leptonic decays of the
$W$ boson to $e$, $\mu$\ and $\tau$ are expected to be equal to very high
precision, given the small sizes of the lepton masses compared to the $W$ boson
mass. This assumption has been tested in the production of $W$-boson pairs in
$e^+e^-$ collisions at LEP2, in the production of single $W$ bosons at the
Tevatron and Large Hadron Collider (LHC), and by exploiting the two $W$ bosons
produced in \ttbar\ events at the LHC. The most precise measurement of \rwme,
the ratio of branching ratios for $W\rightarrow\mu\nu$ and
$W\rightarrow e\nu$, was performed by the CMS collaboration with
$pp\rightarrow\ttbar$ events at \sxyt. This analysis made use of
a global fit to lepton and jet multiplicities, together with kinematic variables
and the identification of jets originating from $b$-quarks,
and achieved a precision of 0.9\%~\cite{CMS-SMP-18-011}.
Measurements of $pp\rightarrow W$ cross-sections in the $W\rightarrow e\nu$ and
$W\rightarrow\mu\nu$ decay channels from the
ATLAS and LHCb experiments~\cite{STDM-2012-20,lhcbwbr},
and measurements in $e^+e^-\rightarrow W^+W^-$ events
from the ALEPH, DELPHI, L3 and OPAL experiments
at LEP2~\cite{lepwbr} also contribute significantly to the combined value of
$\rwme=1.002\pm 0.006$ determined by the Particle Data Group~\cite{pdg}.
 
This paper describes a measurement of \rwme\ using $W$ bosons produced from the
decay of top quarks in $pp\rightarrow\ttbar$ events selected from the full
Run~2 ATLAS $pp$ collision data sample at \sxyt. Final states
with two opposite-charge leptons (electrons or muons, $\ell=e$ or $\mu$) and
one or two jets tagged as likely to contain $b$-hadrons are selected,
allowing \rwme\ to be derived
from a comparison of the \ttbar\ production cross-section measured
in the $ee$, $e\mu$ and $\mu\mu$ channels. Many systematic uncertainties
related to \ttbar\ and background physics modelling cancel in this direct
measurement of \rwme, but it is still limited by uncertainties related to the
identification of electrons and muons. The latter can be reduced by
making a simultaneous measurement of the analogous ratio \rzme\ for $Z$ bosons, i.e.\ the ratio
of branching ratios for \zmm\ and \zee, using inclusive \zll\
events in the same data sample. The main measured parameter of interest
becomes $\rwzme=\rwme/\sqrt{\rzme}$, and the final result is then
obtained from \rwzme\ and the precise measurement of $\rzme=1.0009\pm 0.0028$
from the LEP and SLD experiments~\cite{lepzew,pdg}, taken as an
external input parameter. The \ttbar\ and \zll\
cross-sections, \xtt\ and \xz, are also measured as by-products of this
procedure. The value of \xtt\ is defined inclusively with respect
to all \ttbar\ final states, whereas \xz\ is defined for decays into a single
dilepton flavour $\ell\ell$.
 
The data and samples of Monte Carlo simulated events used in this analysis are
described in Section~\ref{s:datmc}, followed by the event reconstruction
and selection in Section~\ref{s:evtsel}. The analysis method is described
in Section~\ref{s:meth}, and supporting measurements of lepton isolation
efficiencies are outlined in Section~\ref{s:isol}. Systematic uncertainties
are detailed in Section~\ref{s:syst} and the results in Section~\ref{s:res}.
Finally, the conclusion is given in Section~\ref{s:conc}.
 
\section{Data and simulated event samples}\label{s:datmc}
 
The ATLAS detector~\cite{PERF-2007-01,PIX-2018-001,ATLAS-TDR-19} at the LHC
covers nearly the entire solid angle around the collision point.
It consists of an inner tracking
detector surrounded by a thin superconducting solenoid producing
a 2T axial magnetic field, electromagnetic and hadronic calorimeters,
and an external muon spectrometer incorporating three large toroidal
magnet assemblies.
The analysis was performed on samples of proton--proton collision data
collected at \sxyt\ in 2015--18, corresponding to an integrated
luminosity of $\intlumia\pm\intlumie$\,\ifb\
after data quality requirements~\cite{DAPR-2021-01,DAPR-2018-01}.
Events were required to pass a single-electron or single-muon
trigger~\cite{TRIG-2018-05,TRIG-2018-01}, with transverse momentum ($\pt$)
thresholds that
were progressively raised during the data-taking period.\footnote{
ATLAS uses a right-handed coordinate system with its origin at
the nominal interaction point in the centre of the detector, and the $z$-axis
along the beam line. Pseudorapidity is defined in terms of the polar angle
$\theta$ as $\eta=-\ln\tan{\theta/2}$, and transverse momentum is
defined relative to the beam line as $\pt=p\sin\theta$.
The azimuthal angle around the beam line is denoted by
$\phi$, and distances in $(\eta,\phi)$ space by
$\Delta R=\sqrt{(\Delta\eta)^2+(\Delta\phi)^2}$.}
The electron trigger reached the efficiency plateau region for electrons with
reconstructed $\pt>25$\,\GeV\ in 2015 and for $\pt>27$\,\GeV for 2016--18,
the corresponding thresholds for the muon trigger being 21\,\GeV\ for
2015 and 27.3\,\GeV thereafter. Each triggered event also includes the
signals from on average~33 superimposed inelastic $pp$ collisions,
referred to as pileup.
 
Monte Carlo simulated event samples were used to develop the analysis
procedures, to evaluate signal and background contributions, and to compare
with data. Samples were processed using either the full ATLAS detector
simulation~\cite{SOFT-2010-01} based on {\scr Geant4}~\cite{geant4},
or with a faster simulation making use of parameterised showers in the
calorimeters~\cite{ATL-PHYS-PUB-2010-013}. The effects of pileup were
simulated by generating additional inelastic $pp$ collisions with
{\scr Pythia8} (v8.186)~\cite{pythia8} using the A3 set of parameter values
(tune)~\cite{ATL-PHYS-PUB-2016-017} and overlaying them on the primary
simulated events, so as to match the distribution of the number of inelastic
events per bunch crossing observed in the data. These combined
events were then processed using the same reconstruction and analysis
chain as the data~\cite{ATL-SOFT-PUB-2021-001}.
Small corrections were applied to lepton and jet energy
scales~\cite{EGAM-2018-01,MUON-2022-01,JETM-2018-05}, and to lepton and
$b$-tagging efficiencies~\cite{EGAM-2021-01,MUON-2018-03,FTAG-2018-01},
in order to improve agreement with the response
observed in data. Further topology-specific lepton isolation corrections
were applied as discussed in Section~\ref{s:isol}.
 
The baseline simulated \ttbar\ sample was produced with the
{\scr Powheg-Box} v2 event generator~\cite{powheg,powheg2,powheg3,powheghvq}
(referred to hereafter as {\scr Powheg}), which
implements matrix-elements at next-to-leading-order (NLO) in the strong
coupling constant \alphas, using the NNPDF3.0 NLO parton distribution function
(PDF) set~\cite{nnpdf3}. The parton shower, hadronisation and underlying
event modelling was performed using {\scr Pythia8}
(v8.210) with the NNPDF2.3 LO PDF set~\cite{nnpdf23}, the A14
tune~\cite{ATL-PHYS-PUB-2014-021}, and additional parameters configured as
described in Ref.~\cite{ATL-PHYS-PUB-2016-020}. Modelling uncertainties
were assessed by using alternative samples generated using
{\scr MadGraph5\_aMC@NLO}
(referred to hereafter as {\scr aMC@NLO})~\cite{amcnlo}
interfaced to {\scr Pythia8}, and {\scr Powheg} interfaced
to {\scr Herwig7.1}~\cite{herwigpp,herwig71}, as discussed in
Ref.~\cite{ATL-PHYS-PUB-2020-023}. Further variations were
obtained from the baseline {\scr Powheg\,+\,Pythia8} sample,
by using event weights to change the quantum chromodynamics (QCD)
factorisation and renormalisation scales, and the amounts of initial and
final state radiation. The top quark mass was set to $\mtop=172.5$\,\GeV,
the $W\rightarrow\ell\nu$ branching
ratio to the Standard Model prediction of 0.1082 for each lepton flavour
($e$, $\mu$ and $\tau$)~\cite{wpdg},
and {\scr EvtGen}~\cite{evtgen} was used to handle the decays
of $b$- and $c$-flavoured hadrons. The \ttbar\ samples
were normalised to a reference cross-section of $832\pm 35^{+20}_{-29}$\,pb,
where the first uncertainty corresponds to PDF uncertainties and the second to
QCD scale uncertainties. This value was
calculated at next-to-next-to-leading-order (NNLO)
accuracy in \alphas, including the resummation of
next-to-next-to-leading-logarithmic (NNLL) soft gluon terms~\cite{topxtheo5},
using {\scr Top++ 2.0}~\cite{toppp} as described in Ref.~\cite{TOPQ-2018-17}.
The associated production of a $W$ boson and a top quark ($Wt$) is a background
in the \ttbar\ cross-section measurement but contributes sensitivity to
\rwme, as it gives rise to final states with two real $W$ bosons. It was
simulated with {\scr Powheg\,+\,Pythia8} with the same setup as for the \ttbar\
sample. The interference between the \ttbar\ and $Wt$ amplitudes was
modelled using the diagram removal scheme~\cite{wtinter1,wtinter2}.
The $Wt$ cross-section was taken to be $79.3\pm 2.2$ (PDF) $^{+1.9}_{-1.8}$
(QCD scale)\,pb, based on an NLO calculation with the addition of third-order
corrections resumming NNLL soft gluon contributions~\cite{kidonakis21}.
 
The dilepton plus $b$-tagged
jet signature can also arise from $Z$-boson production with additional
jets. In the \ttbar\ measurement, this background was modelled using
{\scr Sherpa 2.2.11}~\cite{sherpa22} for $Z\rightarrow ee/\mu\mu$ and
{\scr Sherpa 2.2.14} for \ztt, with NLO matrix elements for up to
two partons, and leading-order matrix elements for up to five partons,
calculated with the {\scr Comix}~\cite{comix} and
{\scr OpenLoops}~\cite{openloops} libraries and matched with the
{\scr Sherpa} parton shower~\cite{sherpaps} using the {\scr MEPS@NLO}
prescription~\cite{mepsnlo1,mepsnlo2,mepsnlo3,mepsnlo4}.
The samples include the
off-shell $Z/\gamma^*$ and interference contribution and have dilepton
invariant mass $\mll>10$\,\GeV. They were generated using the
NNPDF3.0 PDF set and normalised to an NNLO cross-section
prediction~\cite{zjnnlo}.
For the inclusive \zll\ selections used in the normalisation
measurement of \rzme, where jets are less important,
\zll\ events were modelled using {\scr Powheg} v1~\cite{powhegv}
interfaced to {\scr Pythia8} (v8.186) with the AZNLO tune~\cite{STDM-2012-23}
and the CT10 PDF set~\cite{cttenpdf}, including
$Z/\gamma^*$ and interference contributions, and generating events with
dilepton invariant mass $\mll>60$\,\GeV. These events were reweighted to data
as a function of \ptll\ in order to improve the
modelling of the reconstructed $Z$-boson transverse momentum spectrum,
and the samples were normalised to a reference cross-section
of $\xz=1951$\,pb, based on predictions from {\scr Fewz}~\cite{fewz}.
 
Smaller contributions to both selections arise from diboson production
($WW$, $WZ$ and $ZZ$), which was modelled using {\scr Sherpa 2.2.2},
analogously to $Z$+jets production. Production of \ttbar\ in association
with a leptonically decaying $W$, $Z$ or Higgs boson, or
an additional \ttbar\ pair,  gives a negligible
contribution to the opposite-charge dilepton samples, but is significant
in the same-charge control samples used to assess the background from
misidentified leptons in the \ttbar\ selection. These processes were
modelled at NLO using {\scr Powheg\,+\,Pythia8} or {\scr aMC@NLO\,+\,Pythia8}.
Additional background arises from events where
at least one lepton is not a prompt lepton from a $W$ or $Z$ decay (including
via leptonic $\tau$ decays), but is a misidentified lepton, i.e.\
a non-prompt lepton from the decay of a bottom or charm hadron, an electron
from a photon conversion, a hadronic jet misidentified as an electron,
or a muon produced from the decay in flight of a pion or kaon.
Events with one prompt and one misidentified lepton can arise from
\ttbar\ or $Wt$ events with one hadronically decaying $W$ boson (modelled as
described above), $W$+jets production (modelled with {\scr Sherpa 2.2.1})
or $t$-channel single top quark production (modelled with
{\scr Powheg\,+\,Pythia8}). Processes with two misidentified leptons
(e.g. from inclusive \bbbar\ or \ccbar\ production) are negligible for
the \ttbar\ selection, and the corresponding background in the inclusive
\zll\ selection was modelled from data without relying on the simulation of
such processes (see Section~\ref{s:meth}).

\section{Event reconstruction and selection}\label{s:evtsel}
 
This analysis makes use of reconstructed electrons, muons and $b$-tagged jets.
Electron candidates were reconstructed from a localised cluster of energy
deposits in the
electromagnetic calorimeter matched to a track in the inner detector, passing
the `Medium' likelihood-based requirement of Ref.~\cite{EGAM-2018-01}.
They were required to have transverse momentum $\pt>20$\,\GeV\
and pseudorapidity
$|\eta|<2.47$, excluding the transition region between the barrel and endcap
electromagnetic calorimeters, $1.37<|\eta|<1.52$, and to be consistent with
originating from the signal primary vertex. The latter was defined as the
reconstructed vertex with the highest sum of $\pt^2$ of associated tracks.
To reduce background from
non-prompt electrons, electron candidates were further required
to pass the `Tight' isolation requirements of Ref.~\cite{EGAM-2018-01},
based on the amount of summed calorimeter energy and track transverse momentum
close to the electron. Muon candidates were reconstructed by combining
tracks from the inner detector with matching tracks reconstructed in the
muon spectrometer, and were required to
have $\pt>20$\,\GeV, $|\eta|<2.5$ and to satisfy the `Medium' requirements
of Ref.~\cite{MUON-2018-03}. Muons were also required to be consistent
with the signal primary vertex and to satisfy the `Tight' isolation requirements
of Ref.~\cite{MUON-2018-03}.
 
Jets were reconstructed using the anti-$k_t$ algorithm~\cite{antikt,fastjet}
with radius parameter $R=0.4$, starting from particle-flow objects that combine
information from topological clusters of calorimeter energy deposits and
inner-detector tracks~\cite{PERF-2015-09}.
After calibration using information from
both simulation and data~\cite{JETM-2018-05}, jets were required to have
$\pt>25$\,\GeV\ and $|\eta|<2.5$, and jets with $\pt<60$\,\GeV\
and $|\eta|<2.4$ were subject to additional pileup rejection criteria
using the multivariate jet-vertex tagger (JVT)~\cite{PERF-2014-03}.
To prevent double counting of electron energy deposits as jets, the closest
jet to an electron candidate was removed if it was within $\Delta R=0.2$
of the electron. Furthermore, to reduce the contribution of leptons
from heavy-flavour hadron decays inside jets, leptons within $\Delta R=0.4$
of selected jets were discarded, unless the lepton was a muon and the jet
had fewer than three associated tracks, in which case the jet was discarded.
Jets likely to contain $b$-hadrons were tagged using the DL1r
algorithm~\cite{FTAG-2018-01,FTAG-2019-07}, a multivariate discriminant based
on deep-learning techniques making use of track impact parameters and
reconstructed secondary vertices. A tagger working point with an efficiency
of 70\% for tagging $b$-quark jets from top-quark decays in simulated
\ttbar\ events was used, corresponding to rejection factors of about~380
against light quark and gluon jets, and~10 against jets originating
from charm quarks.
 
Selected events were required to have exactly two leptons (electrons or muons)
passing the requirements above, of opposite charges, with at least one of the
leptons being matched to a corresponding electron or muon trigger signature.
The leptons were additionally
required to have $\pt>27.3$\,\GeV, to further reduce the contribution
of events with misidentified leptons, and to ensure all leptons were above
the muon trigger \pt\ threshold. For the \ttbar\ selection ($ee$, $e\mu$
and $\mu\mu$), events were additionally required to satisfy $\mll>30$\,\GeV\
and to have exactly one or exactly two $b$-tagged jets with $\pt>30$\,\GeV.
For the inclusive \zll\ selection, only $ee$ and $\mu\mu$ events were
retained, with $66\,\GeV<\mll<116$\,\GeV, and no requirements were made on
jet or $b$-tagged jet multiplicity. Table~\ref{t:evtselreq} summarises
the event selection.
 
\begin{table}[tb]
\caption{\label{t:evtselreq}Summary of the common object selection,
and event selections for \ttbar\ and $Z$ final states.}
\centering
 
\begin{tabular}{l|c|c}\hline
Object selection & \multicolumn{2}{l}{} \\
\hline
Electrons & \multicolumn{2}{l}{$\pt>27.3$\,\GeV, $|\eta|<1.37$ or $1.52<|\eta|<2.47$} \\
Muons & \multicolumn{2}{l}{$\pt>27.3$\,\GeV, $|\eta|<2.5$} \\
$b$-tagged jets & \multicolumn{2}{l}{$\pt>30.0\,\GeV$, $|\eta|<2.5$, $b$-tagging DL1r 70\%} \\
\hline
Event selection & \hspace{7mm} $\ttbar\rightarrow\ell\ell\bbbar\nu\bar{\nu}$ \hspace{7mm}\rule[-1mm]{0mm}{5mm} & \zll\ \\
\hline
Dilepton flavour ($\ell^+\ell^-$) & $ee$, $e\mu$, $\mu\mu$ & $ee$, $\mu\mu$ \\
Dilepton invariant mass & $\mll>30$\,\GeV & $66\,\GeV<\mll<116$\,\GeV \\
$b$-tagged jet multiplicity & 1 or 2 & -- \\
\hline
\end{tabular}
\end{table}
 
The numbers of leptons in simulated \ttbar\ events passing the $ee$ and $\mu\mu$
\ttbar\ selections are shown as the points and dashed red lines in
Figure~\ref{f:leptw}(a). The distributions are different, due mainly to the
smaller electron efficiency in the forward region at high $|\eta|$, the
gap in electron acceptance at $1.37<|\eta|<1.52$, and the reduction in
muon acceptance at $\eta\approx 0$ due to detector services. The electron
and muon efficiencies also evolve differently with \pt. To minimise physics
modelling uncertainties in the measurement of \rwme, it is important that
the kinematic dependencies of the electron and muon identification efficiencies
are as similar as possible. This was achieved by applying an $\eta$- and
\pt-dependent weight to each muon, as shown in Figure~\ref{f:leptw}(b).
The weights were derived so as to equalise the two-dimensional distributions of
lepton \pt\ and $\eta$ in simulated $ee$ and $\mu\mu$ \ttbar\ events,
and normalised
so that the total number of selected events is similar in the two channels.
The effect on the muon $\eta$ distribution is shown by the blue solid line
in Figure~\ref{f:leptw}(a). The muon efficiency loss at $\eta\approx 0$ was
compensated over a wide region corresponding to a single bin with $|\eta|<0.5$
to avoid giving muons at $\eta\approx 0$ large weights,
but the physics modelling effects in this region are small. The muon efficiency
weighting affects the event counts in the $\ttbar\rightarrow\mu\mu$ and $e\mu$
selections, and the inclusive \zmm\ selection. It was applied to both data
and simulation, and is included in all distributions, event counts and
efficiencies shown in this paper.

\begin{figure}[tp]
\parbox{83mm}{\includegraphics[width=74mm]{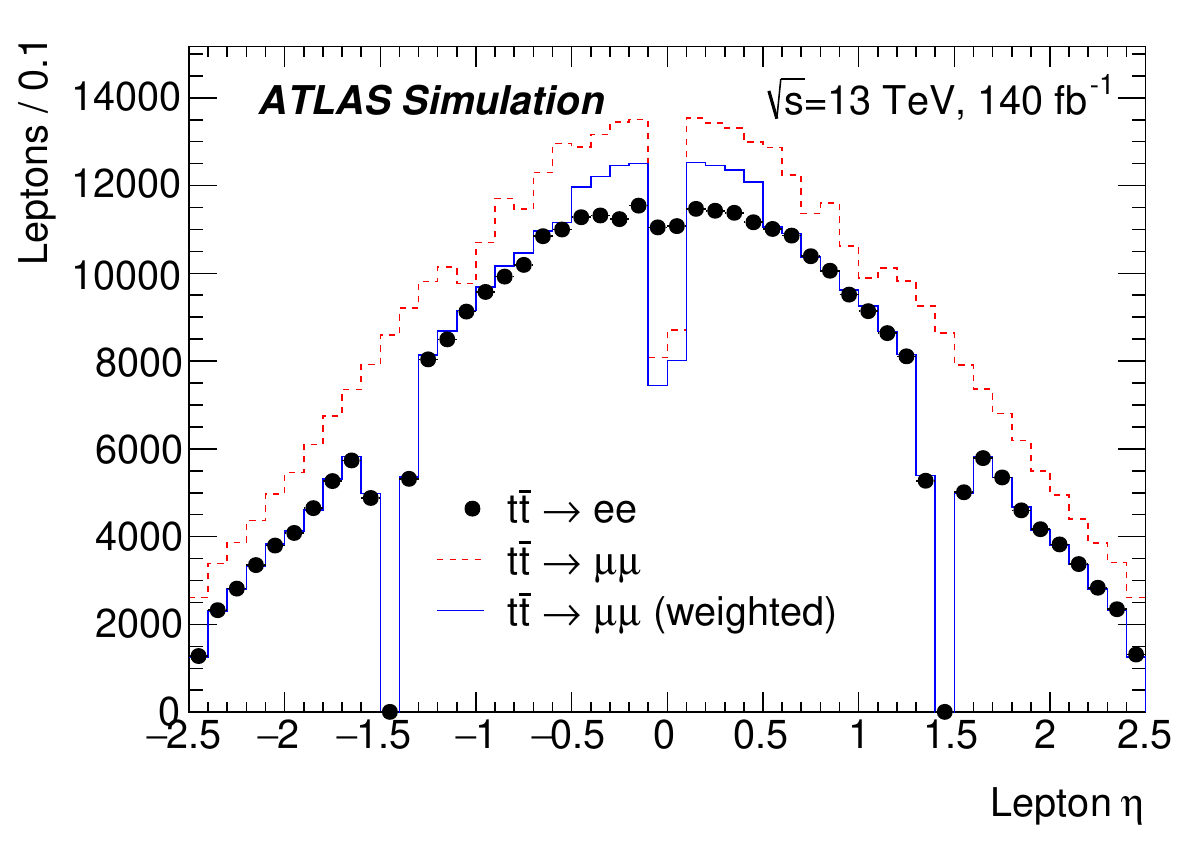}\vspace{-7mm}\center{(a)}}
\parbox{83mm}{\includegraphics[width=74mm]{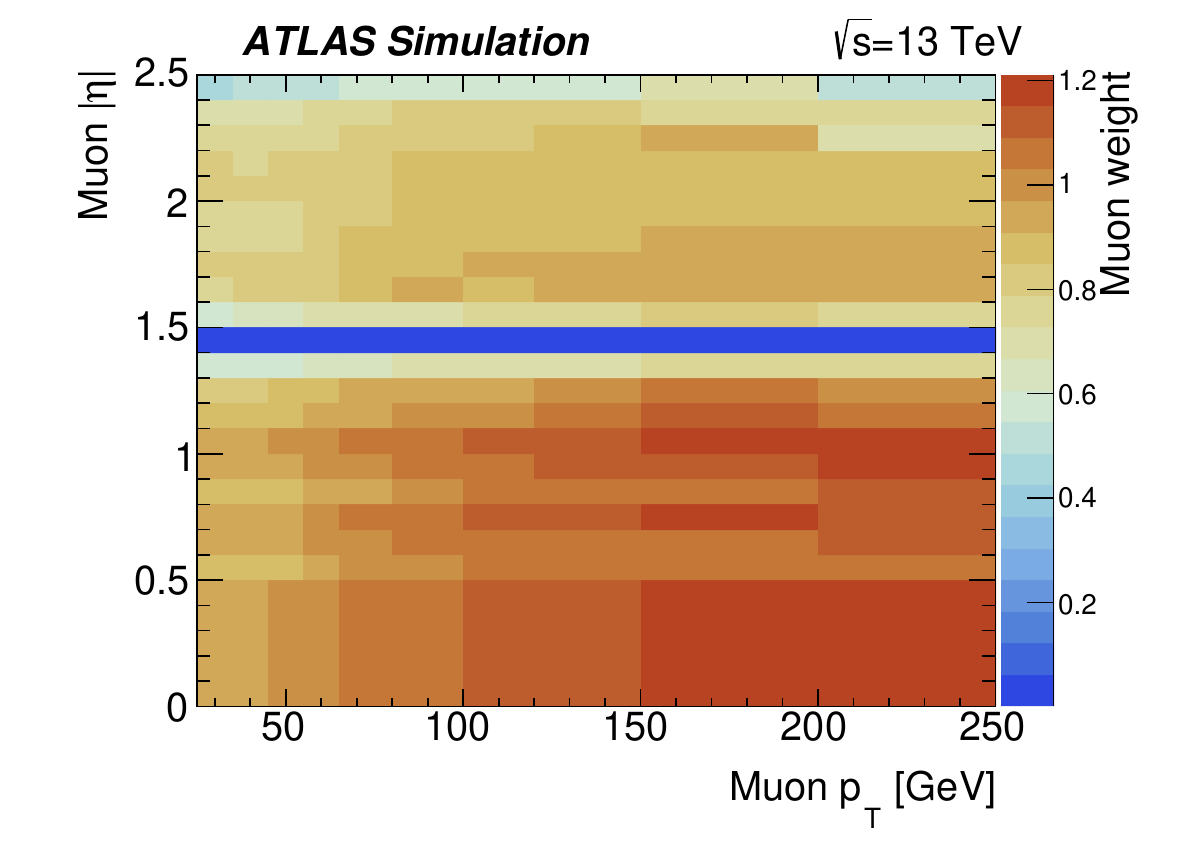}\vspace{-7mm}\center{(b)}}
\caption{\label{f:leptw}(a) Number of selected leptons as a function of $\eta$
in simulated \ttbar\ events with at least one $b$-tagged jet in the
$ee$ (points) and $\mu\mu$ (red dashed line)
channels, and number of selected leptons in the $\mu\mu$ channel after muon
weighting (blue solid line); (b) muon efficiency weights as a function
of $|\eta|$ and \pt.}
\end{figure}
 
Figure~\ref{f:dmcsf} shows comparisons of data and simulation for selected
events in the $ee$ (left column) and $\mu\mu$ (right column) channels,
additionally requiring $|\mll-m_Z|>10$\,\GeV\ to reduce the $Z$+jets
contribution.
The simulation prediction uses the reference values for the \ttbar\ and $Z$
cross-sections given in Section~\ref{s:datmc} and assumes $\rwme=\rzme=1$.
The baseline prediction shown by the black histogram uses
{\scr Powheg\,+\,Pythia8} \ttbar\ events, which are known not to
reproduce the top quark \pt\ spectrum measured in data~\cite{TOPQ-2016-01}
or predicted by NNLO calculations~\cite{topptnnlo}.
The red dotted line shows the prediction from {\scr Powheg\,+\,Pythia8}
reweighted using a linear function of top quark \pt\ as  discussed in
Ref.~\cite{TOPQ-2018-17} in order to better describe the measurement of
Ref.~\cite{TOPQ-2016-01}. The green dashed line shows the prediction using
\ttbar\ events generated with {\scr Powheg\,+\,Herwig7}.
\begin{figure}[tp]
\parbox{83mm}{\includegraphics[width=73mm]{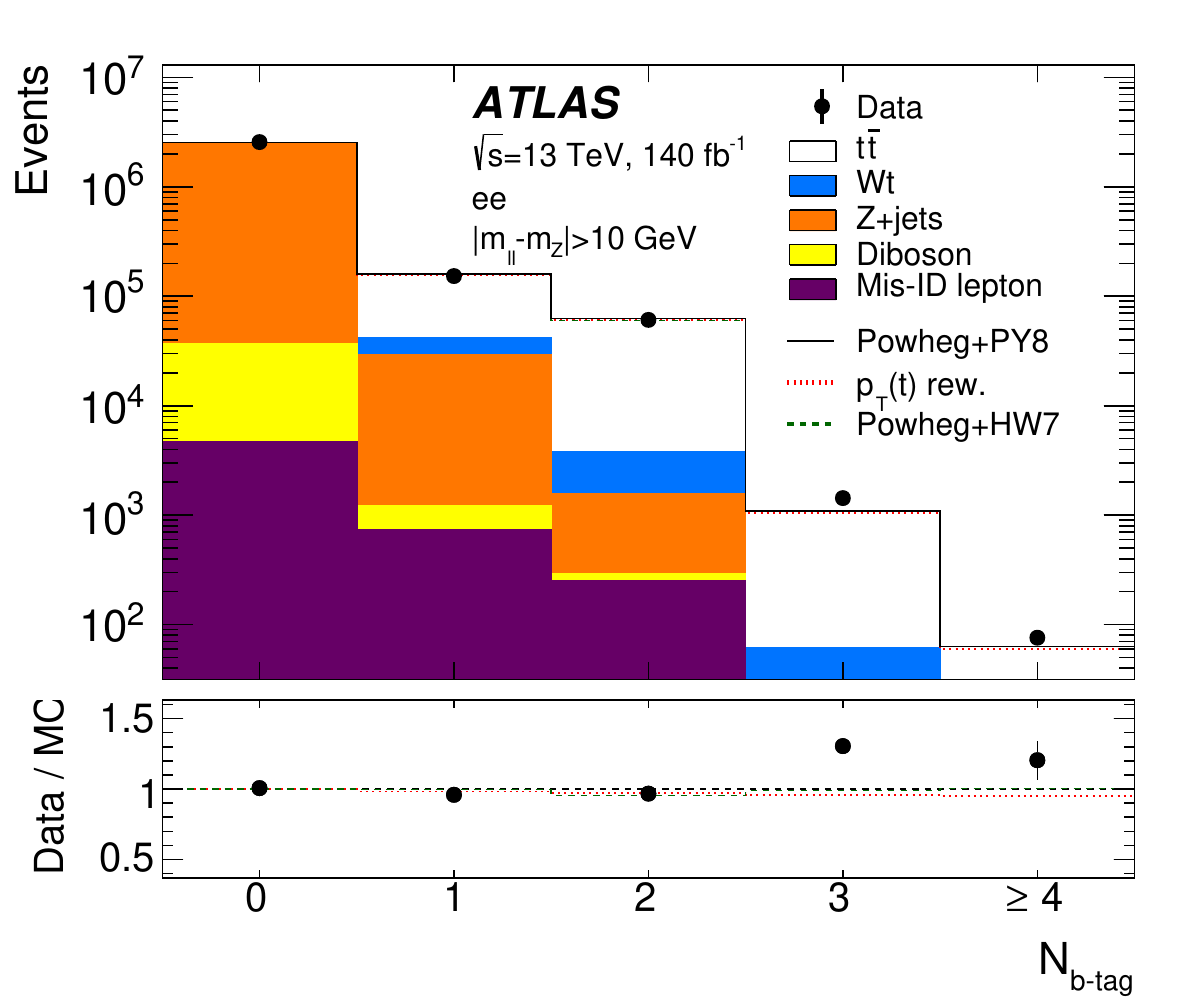}\vspace{-7mm}\center{(a)}}
\parbox{83mm}{\includegraphics[width=73mm]{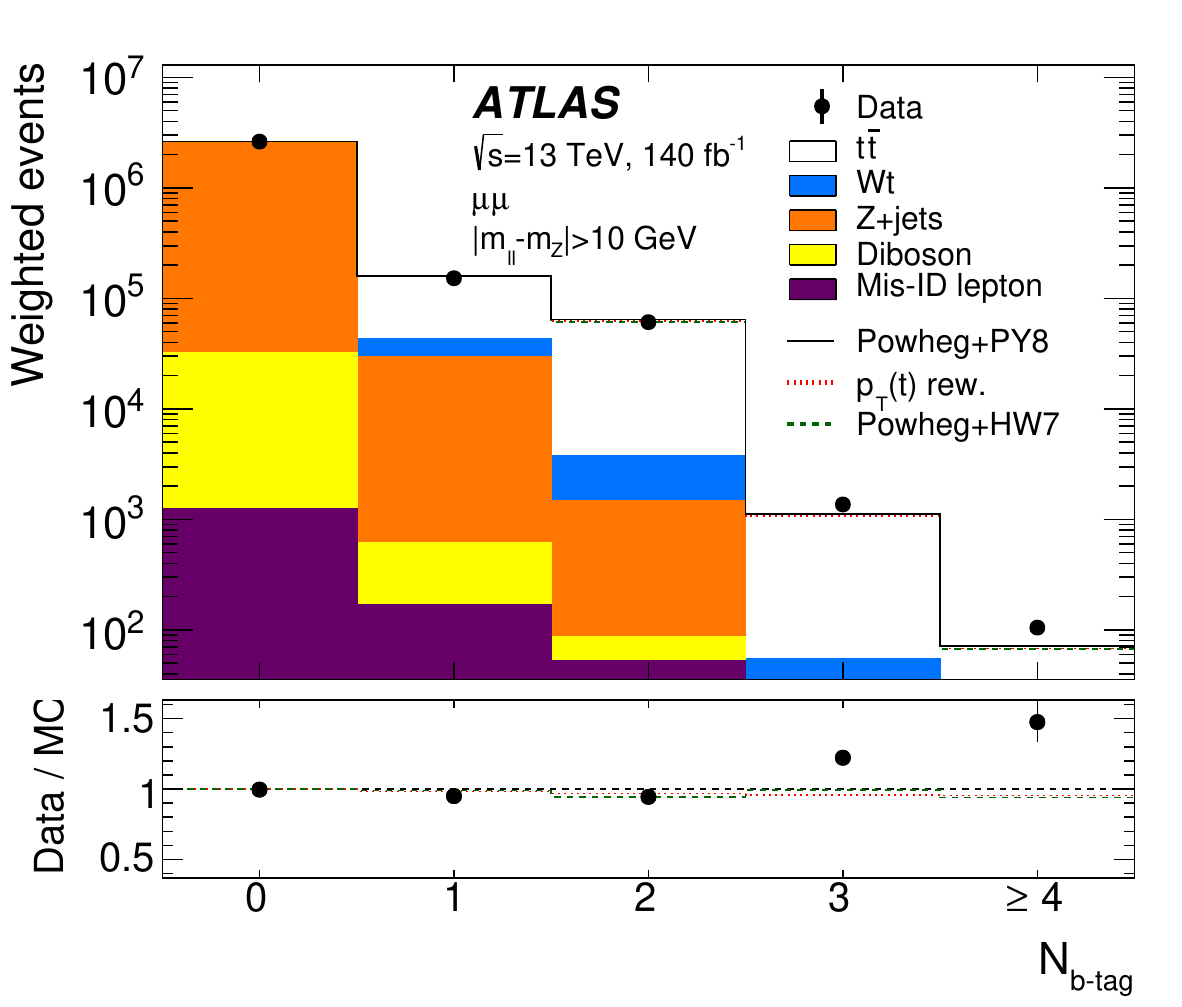}\vspace{-7mm}\center{(b)}
}
\parbox{83mm}{\includegraphics[width=73mm]{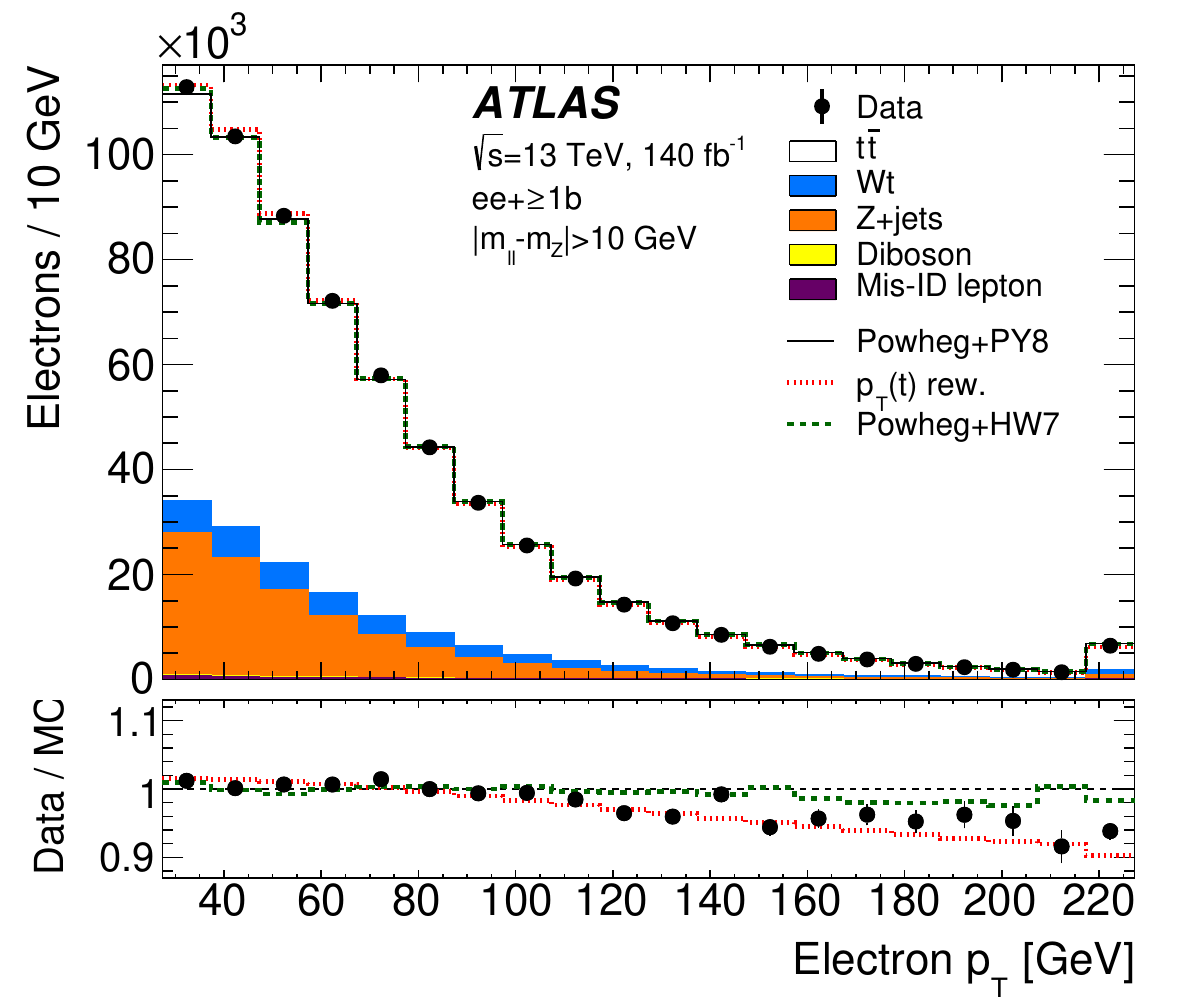}\vspace{-7mm}\center{(c)}}
\parbox{83mm}{\includegraphics[width=73mm]{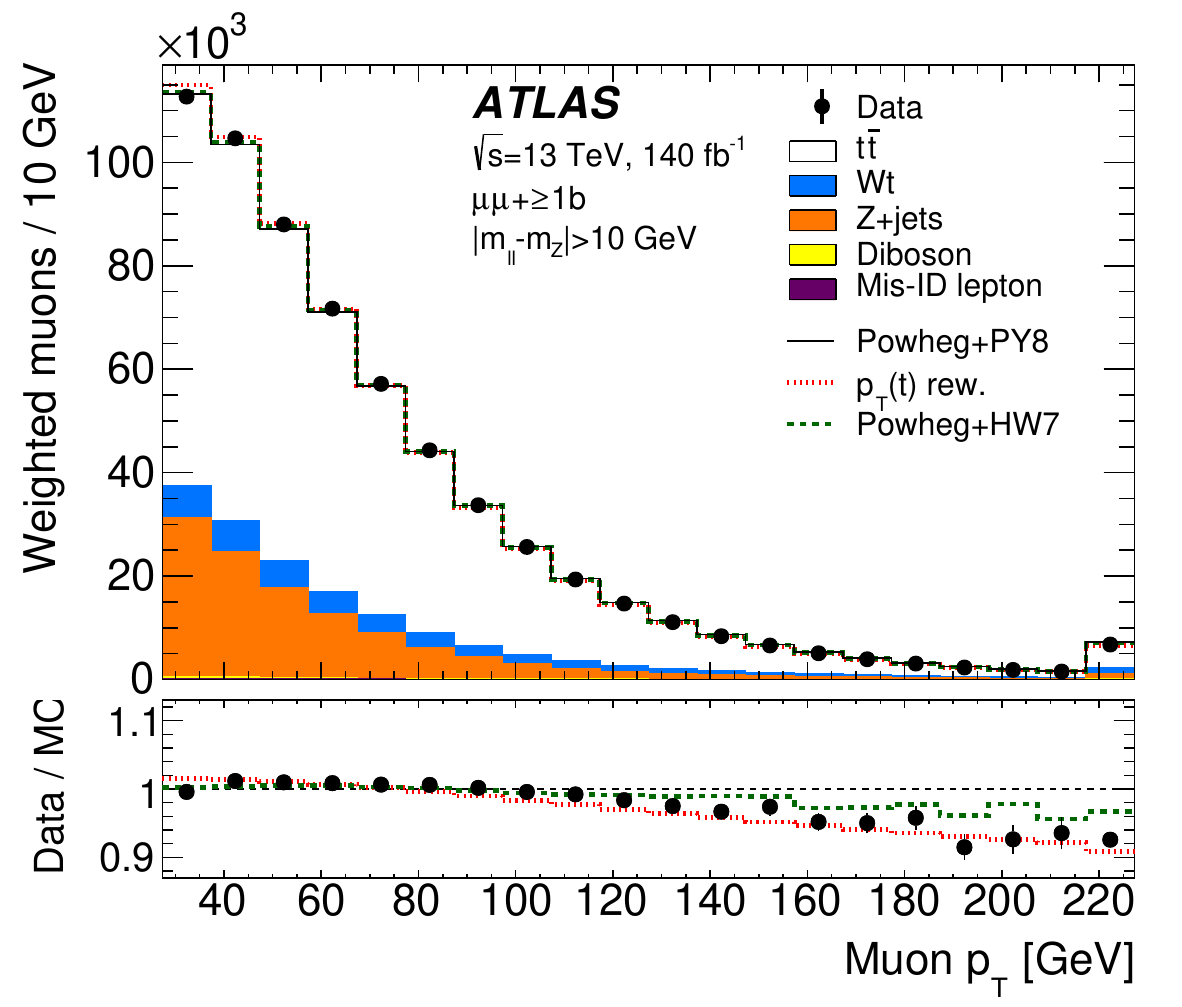}\vspace{-7mm}\center{(d)}
}
\parbox{83mm}{\includegraphics[width=73mm]{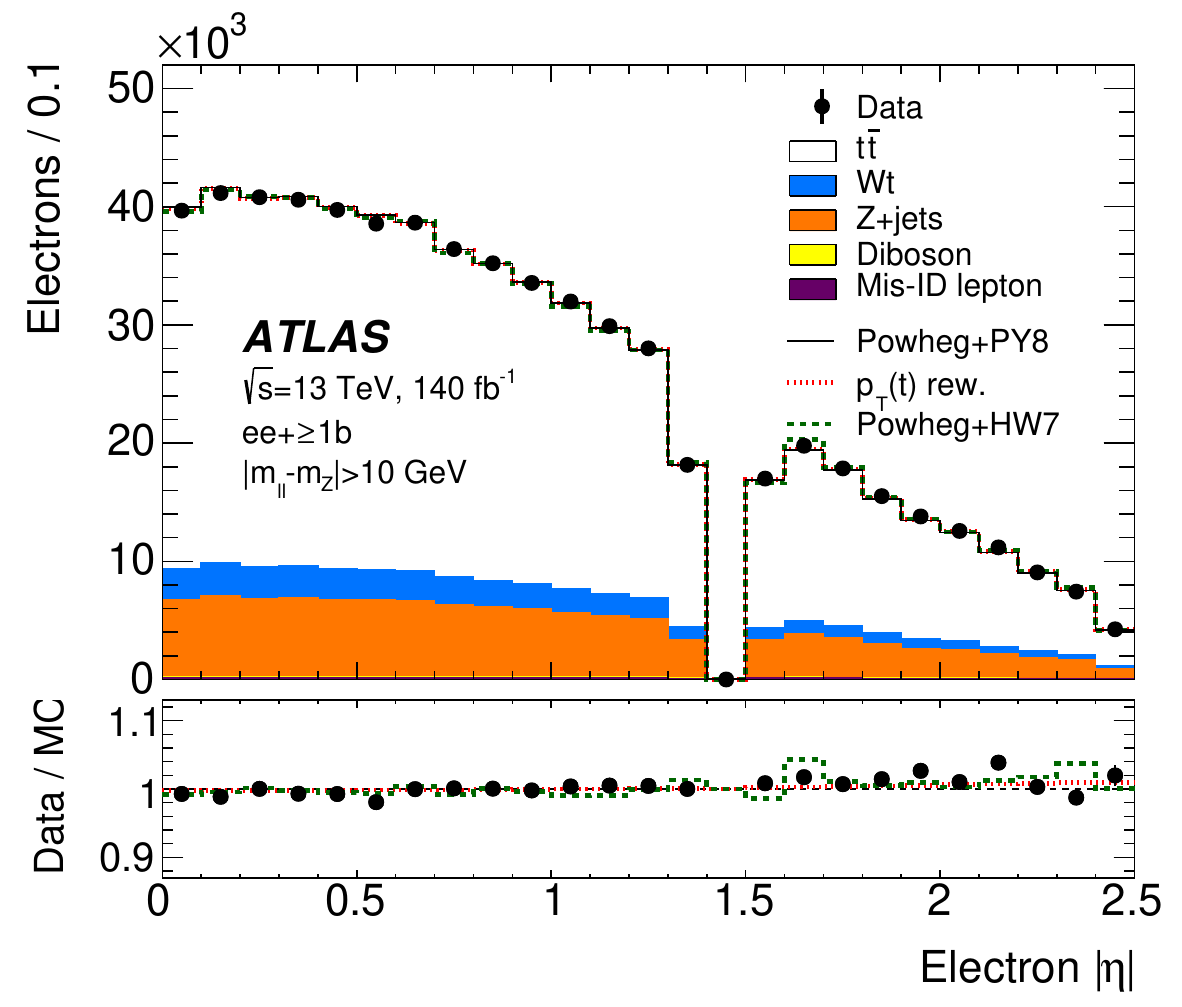}\vspace{-7mm}\center{(e)}}
\parbox{83mm}{\includegraphics[width=73mm]{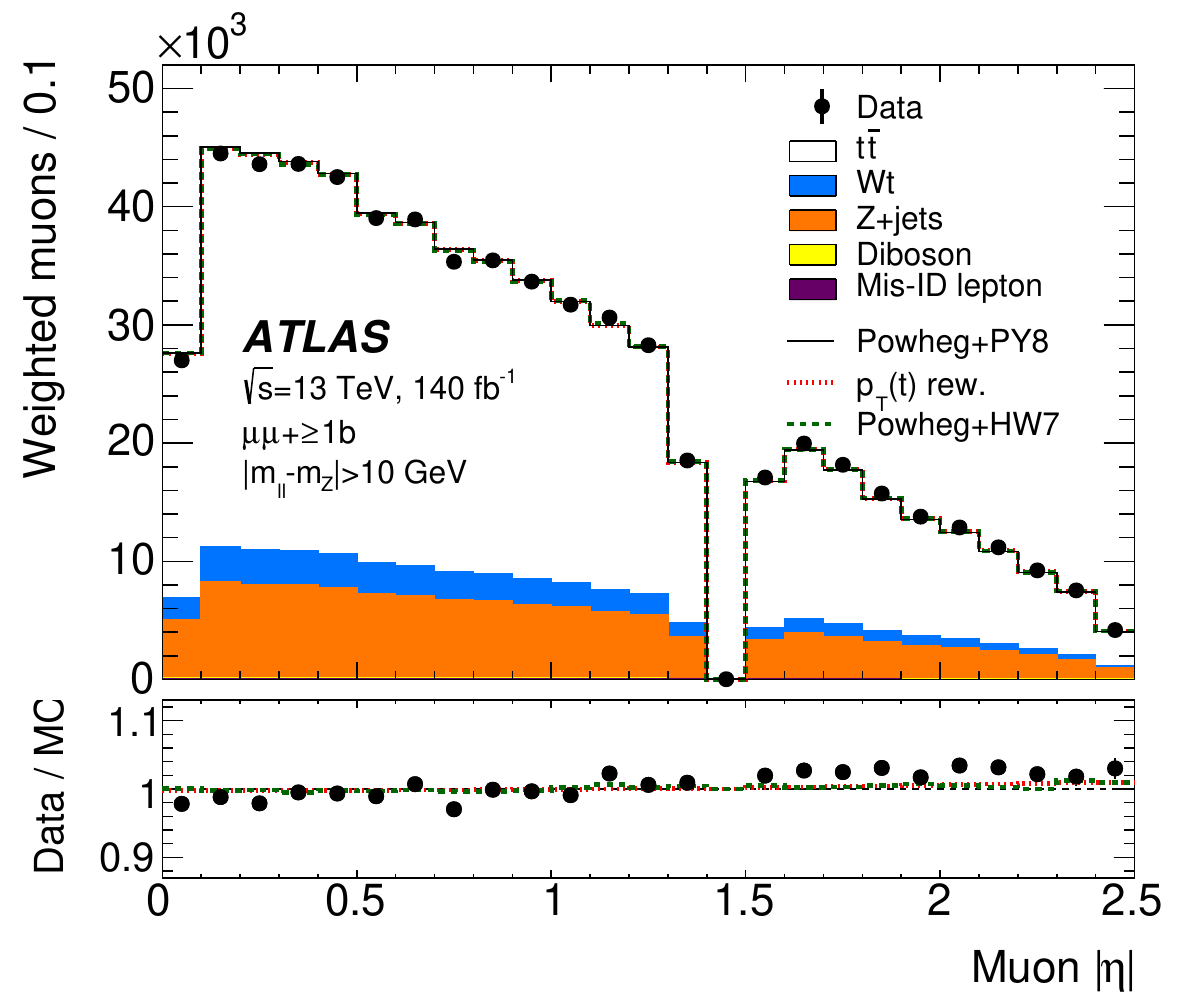}\vspace{-7mm}\center{(f)}
}
\caption{\label{f:dmcsf}Distributions of (a, b) the number of $b$-tagged jets
in selected opposite-sign dilepton events with $|\mll-m_Z|>10$\,\GeV\
(without applying the $b$-tagged jet multiplicity requirement), together with
(c, d) the lepton transverse momentum and (e, f) the lepton pseudorapidity
in such events with at least one $b$-tagged jet, showing $ee$ (left column)
and weighted $\mu\mu$ (right column) events separately. The data is shown
by the points with statistical error bars, compared to the prediction from
simulation normalised to the data integrated luminosity in (a, b) and to the
same number of selected events in (c--f). The predicted contributions
from \ttbar, $Wt$, $Z$+jets, dibosons and events with misidentified leptons
are shown separately. The red dotted line shows the prediction with the
reweighted top quark \pt\ distribution, and the green dashed line that with
{\scr Powheg\,+\,Herwig7} \ttbar\ events instead of {\scr Powheg\,+\,Pythia8}.
The lower plots show the
ratio of data to the baseline simulation, and the ratios of the alternative
simulation predictions to the baseline.
The last bins include the overflows in plots (a--d).}
\end{figure}
Figure~\ref{f:dmcsf}(a--b) shows the multiplicity of $b$-tagged jets, \nbtag,
with the simulation normalised to the same integrated luminosity as the data;
the simulation describes the data well for $\nbtag\leq 2$, but shows
a deficit for $\nbtag\geq 3$. Figure~\ref{f:dmcsf}(c--f) shows the lepton \pt\
and $|\eta|$ distributions for the \ttbar-dominated samples with
at least one $b$-tagged jet, normalising the simulation to the same number
of selected events as the data to focus on shape comparisons.
The samples with {\scr Powheg\,+\,Pythia8} and {\scr Powheg\,+\,Herwig7}
\ttbar\ events both
predict a harder \pt\ spectrum than the data, but the top-quark \pt-reweighted
sample agrees well, as also seen in  Ref.~\cite{TOPQ-2018-17}.
Similar features are seen in the $e\mu$ selection.
The lepton \pt\ and $|\eta|$ distributions in the $ee$ and $\mu\mu$ samples are
similar (apart from at $\eta\approx 0$), demonstrating the effect of the
muon efficiency weighting described above.

\section{Analysis method}\label{s:meth}
 
The \ttbar\ cross-section was measured in each dilepton channel by fitting
the numbers of selected events with one or two $b$-tagged jets to predictions
based on the assumed \ttbar\ cross-section, leptonic selection efficiencies
\epsll\ and estimated non-\ttbar\ background. In the same-flavour channels, the
dilepton invariant mass \mll\ was also exploited to separate signal events
from the dominant $Z$+jets background. This method allows the efficiency
\epsbllp\ for reconstructing and $b$-tagging a $b$-jet from the top quark decay
to be determined from the data
(separately for $\ell\ell'=ee$, $e\mu$ and $\mu\mu$),
and minimises selection efficiency uncertainties due to the
modelling of additional jets from QCD radiation in the \ttbar\ events.
 
Following Ref.~\cite{TOPQ-2018-17}, the inclusive \ttbar\ cross-section was
determined in the $e\mu$ channel from the number of opposite-charge events
with one ($N_1^{e\mu}$) or two ($N_2^{e\mu}$) $b$-tagged jets.
The two event counts satisfy the tagging equations
\begin{equation}
\begin{array}{lllll}
N_1^{e\mu} & = & L \xtt\ \epsem \gemtt\, 2\epsbem (1-\cbem\epsbem) & + & \displaystyle\sum\limits_{k=\mathrm{bkg}}
s_1^k \gemk\, N_1^{e\mu,k}\ \textrm{and}  \\
N_2^{e\mu} & = & L \xtt\ \epsem \gemtt\, \cbem(\epsbem)^2 & + & \displaystyle\sum\limits_{k=\mathrm{bkg}} s_2^k \gemk\, N_2^{e\mu,k},
\end{array}\label{e:tags}
\end{equation}
where $L$ is the integrated luminosity of the sample, \epsem\ is the efficiency
for a \ttbar\ event to pass the opposite-charge $e\mu$ selection
(including the simulated values of the $W\rightarrow\ell\nu$ branching ratios),
and \gemtt\ expresses possible deviations of these branching ratios
from their simulated values. The parameter $\cbem\approx 1$ is a correlation
coefficient that accounts for the fact that the tagging
probabilities of the two $b$-quark jets from the top decays are not completely
independent. It was evaluated
from simulation as $\cbem=4\nttem N^{\ttbar}_2/(N^{\ttbar}_1+2N^{\ttbar}_2)^2$,
where \nttem\ is the number of selected $e\mu$ \ttbar\ events and $N^{\ttbar}_1$
and $N^{\ttbar}_2$ are the numbers of such events with one and two $b$-tagged
jets~\cite{TOPQ-2018-17}.
Background contributions to $N_1$ and $N_2$ were divided into four
sources indexed by $k$: $Wt$, $Z$+jets, dibosons and events with
at least one misidentified lepton. The estimate of each background $k$
was scaled by a factor $s^k_n$ for events with $n=1$ or 2 $b$-tagged jets,
and additionally scaled by \gemk\ to allow for changes in the $W$ or $Z$
leptonic branching ratios.
 
In the same-flavour channels, the events were divided into six bins of
\mll\ to separate the \ttbar\ signal from the large $Z$+jets background.
The bins were
indexed by subscript $m$, with lower bin boundaries at 30, 71, 81, 101,
111 and 151\,\GeV, the last bin including all events with $\mll>151$\,\GeV.
Using the extension of the tagging formalism introduced in
Ref.~\cite{TOPQ-2018-40},
the numbers of opposite-charge $\ell\ell$ events in each bin $m$ with one
and two $b$-tagged jets, $N_{1,m}^{\ell\ell}$ and $N_{2,m}^{\ell\ell}$
can then be expressed as
\begin{equation}
\begin{array}{lllll}
N_{1,m}^{\ell\ell} & = & L \xtt\,\epsll\,\glltt\, 2\epsbll (1-\cbll\epsbll)\,f_{1,m}^{\ell\ell,t\bar{t}} & + &
\displaystyle\sum\limits_{k=\mathrm{bkg}} s_1^k \gllk\, f_{1,m}^{\ell\ell,k} N_{1}^{\ell\ell,k}\ \textrm{and} \\
N_{2,m}^{\ell\ell} & = & L \xtt\,\epsll\,\glltt\,\cbll(\epsbll)^2\,f_{2,m}^{\ell\ell,t\bar{t}} & + &
\displaystyle\sum\limits_{k=\mathrm{bkg}} s_2^k \gllk\, f_{2,m}^{\ell\ell,k} N_{2}^{\ell\ell,k},
\end{array}\label{e:tagsf}
\end{equation}
with separate selection efficiencies \epsll\ and correlation coefficients
\cbll\ for each same-flavour channel ($\ell\ell=ee$ or $\mu\mu$).
The coefficients $f_{1,m}^{\ell\ell,k}$ and $f_{2,m}^{\ell\ell,k}$ describe the
\mll\ distributions, giving the fractions of events that appear in each mass
bin, separately for each dilepton flavour $\ell\ell$,
event source $k$ and $b$-tagged jet multiplicity (1 or~2).
 
This analysis allows the branching ratios  \brwe\ and \brwm\ to differ
via a parameter $\Delta_W$, whilst keeping their average fixed to
$\brmean=0.1082$, the Standard Model prediction used in the simulation.
In this model
\begin{equation}\label{e:rwme}
\rwme = \frac{\brwm}{\brwe} = \frac{\brmean (1+\Delta_W)}{\brmean (1-\Delta_W)}\ ,
\end{equation}
so that $\Delta_W=(\rwme-1)/(\rwme+1)$. The selected \ttbar\ dilepton samples
also include events where one or both leptons arise from a
$W\rightarrow\tau\rightarrow e/\mu$ decay, and the branching ratios
for $W\rightarrow\tau\nu$, $\tau\rightarrow e\nu\bar{\nu}$ and
$\tau\rightarrow \mu\nu\bar{\nu}$ were kept fixed at the values in the
simulation. With these assumptions, the factors \gllptt\ in Eqs.~(\ref{e:tags})
and~(\ref{e:tagsf}) are given by
\begin{equation}\label{e:taufrac}
\begin{array}{llllll}
\geett & = & \ftauee{0}(1-\Delta_W)^2 & +\ftauee{1}(1-\Delta_W) & +\ftauee{2} \rule[-2mm]{0mm}{6mm} \\
\gemtt & = & \ftauem{0}(1-\Delta_W)(1+\Delta_W) & +\ftauem{1} & +\ftauem{2} \rule[-2mm]{0mm}{6mm} \\
\gmmtt & = & \ftaumm{0}(1+\Delta_W)^2 & +\ftaumm{1}(1+\Delta_W) & +\ftaumm{2} \rule[-2mm]{0mm}{6mm} \\
\end{array}\ ,
\end{equation}
where the parameters \ftau{\ell\ell'}{n}\ give the fractions in each
selected dilepton sample where $n$ leptons resulted from
$W\rightarrow\tau\rightarrow e/\mu$ rather than direct
$W\rightarrow e/\mu$ decays. These fractions were taken from simulation,
and are around $\ftau{\ell\ell'}{0}=0.88$, $\ftau{\ell\ell'}{1}=0.11$
and $\ftau{\ell\ell'}{2}=0.004$ for all three dilepton flavour combinations.
Increasing $\ftau{\ell\ell'}{1}$ by 1.3\% and $\ftau{\ell\ell'}{2}$ by 2.6\%,
corresponding to the uncertainty of 1.3\% in $\brwt/\brwm$ measured
in Ref.~\cite{TOPQ-2018-29}, has a negligible effect on the fitted value
of \rwme\ from this analysis.
 
The estimates of the $Wt$ and diboson backgrounds $N_n^{\ell\ell',k}$ in
Eqs.~(\ref{e:tags}) and~(\ref{e:tagsf})
(with $k=Wt$ or diboson) were taken
directly from simulation, with $s_n^k$ fixed to unity. However, since
$Wt$ events have two real $W$ bosons and the diboson background is dominated by
$WW$ production, the corresponding values of \gllpk\ were set equal to \gllptt\
given by Eq.~(\ref{e:taufrac}), effectively treating these backgrounds
as signal for the determination of \rwme. The normalisation factors \rza\ and
\rzb\ for the $Z$+jets background were determined from data, exploiting
the binning of the same-flavour dilepton events in \mll, and applying the
same factors to all three dilepton channels. However, the introduction
of the normalisation measurement of \rzme\ also affects the $Z$+jets
background estimate. Potential deviations of \rzme\ from unity were
described by a parameter $\Delta_Z$, related to \rzme\ by
\begin{equation}\label{e:rzme}
\rzme = \frac{\brzmm}{\brzee} = \frac{\brmeanz (1+\Delta_Z)}{\brmeanz (1-\Delta_Z)}\ ,
\end{equation}
where \brmeanz\ is the average \zll\ branching ratio and
$\Delta_Z=(\rzme-1)/(\rzme+1)$, in analogy to Eq.~(\ref{e:rwme}).
Potential biases in the modelling of the lepton isolation efficiency
in the busy hadronic environment of $Z+b$-jet events
(in particular differences between electrons and muons as discussed in
Section~\ref{s:isol}) were taken into account
by an additional ratio \rzjme\ and associated parameter
$\dzjme=(\rzjme-1)/(\rzjme+1)$.
With these ingredients, the values of \gllpk\ for $Z$+jets events are
given by
\begin{equation}\label{e:zgll}
\begin{array}{lll}
\gxy{ee}{Z+\mathrm{jets}} & = & (1-\Delta_Z)(1-\dzjme) \\
\gxy{e\mu}{Z+\mathrm{jets}} & = & 1 \\
\gxy{\mu\mu}{Z+\mathrm{jets}} & = & (1+\Delta_Z)(1+\dzjme) \\
\end{array}\ .
\end{equation}
The contributions to the backgrounds in Eqs.~(\ref{e:tags}) and~(\ref{e:tagsf})
from events with misidentified leptons were evaluated using a partially
data-driven method, as discussed below.
 
The factors \gllptt\ giving sensitivity to the $W$-boson branching ratios
are related to $\Delta_W$ and hence \rwme\ by Eqs.~(\ref{e:rwme})
and~(\ref{e:taufrac}). However, to reduce sensitivity to uncertainties in
the electron and muon identification efficiencies,
the fit was not performed with
\rwme\ directly, but instead using \rwzme\ and \rzme, where
\begin{equation}\label{e:rwzme}
\rwzme = \frac{\rwme}{\sqrt{\rzme}} = \frac{\brwm}{\brwe} \cdot \sqrt{\frac{\brzee}{\brzmm}} \ .
\end{equation}
The normalisation to $\sqrt{\rzme}$ ensures that the numerator
and denominator of \rwzme\ each contain one power of the electron and muon
efficiencies, reducing the sensitivity of \rwzme\ to uncertainties on these
efficiencies. The value of \rzme\ needed in Eq.~(\ref{e:rwzme}) was determined
from the event counts in the inclusive \zll\ selection, \nzee\ and \nzmm,
given by
\begin{equation}
\begin{array}{lllll}
\nzee & = & L\, \xz\, \epszee (1-\Delta_Z) & + & \displaystyle\sum\limits_{k=\mathrm{bkg}} s_Z^k\, N_Z^{ee,k} \ \textrm{and}\\
\nzmm & = & L\, \xz\, \epszmm (1+\Delta_Z) & + & \displaystyle\sum\limits_{k=\mathrm{bkg}} s_Z^k\, N_Z^{\mu\mu,k} \ ,\\
\end{array}\label{e:ztags}
\end{equation}
where \epszee\ and \epszmm\ are the selection efficiencies in simulation
assuming equal branching ratios for \zee\ and \zmm, and the factors
involving $\Delta_Z$ express the effects of deviations of \rzme\ from
unity. Five sources of backgrounds were considered, indexed by $k$: dibosons,
$Z\rightarrow\tau\tau\rightarrow ee/\mu\mu$, \ttbar, $Wt$ and events
with misidentified leptons. The first four were estimated from simulation,
with the \ttbar\ background being scaled according to the fitted value
of \xtt\ via its normalisation $s_Z^{t\bar{t}}$, and all other $s_Z^k$ values
were fixed to unity. The misidentified-lepton background was estimated
from data as discussed below.
 
All fit parameters were determined simultaneously using a single maximum
likelihood fit to the observed event counts $N_1^{e\mu}$ and $N_2^{e\mu}$ in the
$e\mu$ channel, the observed counts in each dilepton invariant mass bin
$N_{1,m}^{\ell\ell}$ and $N_{2,m}^{\ell\ell}$ for each same-flavour channel,
and the observed counts \nzee\ and \nzmm\ in the inclusive \zll\ selections,
as summarised in Table~\ref{t:fitscheme}.
A Gaussian likelihood formulation was used, taking into account the
probability distributions of the weighted event counts in the $\mu\mu$
and $e\mu$ channels.
The fit has ten free parameters: the four parameters of interest \xtt, \xz,
\rwzme\ and \rzme, the three $b$-tagged jet efficiencies \epsbllp,
the scale factors \rza\ and \rzb\ for the $Z$+jets background,
and the $Z$+jets isolation efficiency parameter \rzjme.
Apart from the integrated luminosity $L$
and the misidentified lepton backgrounds, all other quantities were determined
from simulation, namely the efficiencies \epsllp, \epszee\ and \epszmm,
$b$-tagging correlations \cbllp, $\tau$ fractions $\ftau{\ell\ell'}{n}$,
background counts $N_1^{\ell\ell',k}$, $N_2^{\ell\ell',k}$ and $N_Z^{\ell\ell,k}$,
and mass distributions $f_{1,m}^{\ell\ell,k}$ and $f_{2,m}^{\ell\ell,k}$.
In the baseline simulation, the \ttbar\ dilepton selection efficiencies are
$\epsee=0.316$\%, $\epsem=0.639$\% and $\epsmm=0.314$\%,
the correlation coefficients \cbllp\ are about 1.003 for
all dilepton flavours, and the $Z$ selection efficiencies are
\epszee=16.8\% and \epszmm=17.3\%. The fitted $Z$ cross-section
and \rzme\ are constrained by the inclusive \zll\ selection, \xtt\ is
mainly determined from the $e\mu$ channel, and \rwzme, \rza, \rzb\ and \rzjme\
are mainly determined from the same-flavour \ttbar\ selections.
Around 1\% of the events in the inclusive \zll\ selections are also
included in the same-flavour \ttbar\ selections, but this overlap has
a negligible effect on the analysis.
The analysis procedure was validated using simulation-based pseudo-experiments
with various input values of \xtt, \xz, \rwme\ and \rzme. These tests verified
that the fit gives correct uncertainty estimates and that any residual biases
are much smaller than the data statistical uncertainty.
 
\begin{table}
\caption{\label{t:fitscheme}Summary of the fitted distributions, showing the
distribution variable, number of bins and event count nomenclature for each
sample of selected events.}
\centering
 
\begin{tabular}{l|c|c|c}\hline
Event selection & Variable & Bins & Event count \\
\hline
$e\mu$+1 or 2 $b$-tagged jets & $\nbtag$ & 2 & $N_1^{e\mu}$, $N_2^{e\mu}$ \rule{0mm}{4mm}\\[1mm]
$ee$+1 $b$-tagged jet & \mll & 6 & $N_{1,m}^{ee}$ \\[1mm]
$ee$+2 $b$-tagged jets & \mll & 6 & $N_{2,m}^{ee}$ \\[1mm]
$\mu\mu$+1 $b$-tagged jet & \mll & 6 & $N_{1,m}^{\mu\mu}$ \\[1mm]
$\mu\mu$+2 $b$-tagged jets & \mll & 6 & $N_{2,m}^{\mu\mu}$ \\[1mm]
\zee\ or $\mu\mu$ & channel & 2 & \nzee, \nzmm \\[1mm]
\hline
\end{tabular}
\end{table}
 
The background from events with misidentified leptons in the \ttbar\ selection
was estimated using same-sign~(SS) control samples, selected as described
in Section~\ref{s:evtsel} but requiring two leptons of the same rather than
opposite electric charges. In an extension of the method described in
Ref.~\cite{TOPQ-2018-17}, the misidentified lepton background \nijfake\ in
invariant mass bin $i$ (using a single bin for the $e\mu$ channel) with $j$
$b$-tagged jets, was estimated from the number of SS events in data,
\nijdss, after subtracting the number of prompt SS events \nijphss\ estimated
using simulation, and then scaling by the ratio $R^i_j$ of misidentified-lepton
events in the opposite-sign (OS) and SS samples in simulation:
\begin{eqnarray}
\nijfake & = & R^i_j (\nijdss-\nijphss) \nonumber \ ,\\
R^i_j & = & \frac{\nijfakeos}{\nijfakess} \label{e:fakettr} \ .
\end{eqnarray}
To reduce the pollution of the SS $ee$ and $e\mu$ samples from true OS events
where an electron charge sign was misreconstructed (particularly for \mll\
close to the \zee\ resonance), electrons in the SS sample
were required to be accepted by a charge misidentification boosted decision
tree (BDT)~\cite{EGAM-2018-01} that reduces the rate of electron charge
misidentification by up to an order of magnitude.
The values of $R^i_j$ are sensitive to the composition of the misidentified
lepton background in simulation, and the corresponding uncertainty was
assessed by removing the photon conversion, misidentified
hadron and muon decay in flight contributions in turn, and recalculating
$R^i_j$. The modelling of the charge misidentification BDT was
studied using \zee\ events. An uncertainty
of 25\% on the prompt SS contribution was assumed, covering the uncertainties
in the dominant contributing processes (\ttbar+$W$, $Z$ and $H$, and $WZ$)
and in the rate of electron charge misidentification. The misidentified lepton
background was evaluated to contribute 0.4--1.2\% of the $ee$ opposite-sign
sample (depending on \mll\ and the $b$-tagged jet multiplicity), 0.2--0.8\%
of the $e\mu$ sample and less than 0.4\% of the $\mu\mu$ sample.
The estimates from data are compatible with the predictions from simulation
within the evaluated uncertainties of 30--70\%.
The simulation was also found to provide
a good description of data SS control samples where the isolation requirements
were inverted to increase the misidentified lepton contributions.
 
The misidentified lepton background in the \zll\ selection was estimated from
data in the region $66\,\GeV<\mll<81$\,\GeV\ and $101\,\GeV<\mll<116$\,\GeV\
by defining two orthogonal control samples (B and C) enriched in
misidentified leptons. The B samples were defined by requiring the lower-\pt\
lepton to fail the isolation
requirement, and in the case of electrons also requiring it to fail the Medium
identification requirement but pass a looser requirement. The C samples
were defined by requiring the two leptons to have the same rather than opposite
electric charges. To reduce the background from genuine \zee\ events
where one electron charge sign was mismeasured, both electrons
were required to be accepted by the charge misidentification BDT.
A third control sample (D) was defined by applying both B and C requirements.
Assuming the B and C requirements to be uncorrelated, the number $\nfake{A}$ of
misidentified lepton events in the A (signal) sample was then estimated from
$\nfake{A}=f \nfake{B}\nfake{C}/\nfake{D}$ where $\nfake{X}$ is the number
of observed events in region $X$ ($X=B$, $C$ or $D$) after subtracting the
prompt lepton contribution using simulation, and the factor
$f=(50\,\GeV)/(15\,\GeV+15\,\GeV)=5/3$ linearly interpolates
the estimate over the complete mass range $66\,\GeV<\mll<116$\,\GeV. The
misidentified lepton contributions were found to be $0.39\pm 0.37$\%
of the \zee\ sample and $0.06\pm 0.27$\% of the \zmm\ sample, estimating
the systematic uncertainties by using a tighter anti-isolation requirement
in the B and D samples.

\section{Lepton isolation efficiency measurements}\label{s:isol}
 
The efficiency of the isolation requirements applied to the leptons was
measured directly in data using a tag-and-probe methodology, separately for
the \zll\ and busier $\ttbar\rightarrow\ell\ell$ environments, and
for all leptons with $\pt>20$\,\GeV. In the
\zll\ measurements, opposite-sign $ee$ and $\mu\mu$ pairs were selected,
requiring the tag lepton to satisfy the identification and isolation
cuts described in Section~\ref{s:evtsel} and to be matched to a corresponding
trigger signature. The probe lepton was only required
to satisfy the
identification cuts, and the isolation efficiency was measured from the
fraction of probe leptons with dilepton invariant mass in the range
$80\,\GeV<\mll<102$\,\GeV that pass the isolation requirement, after correcting
for the background from non-prompt leptons, which reaches up to 1\%
in the samples failing the isolation requirement at low lepton \pt.
This background was estimated using a template fit to the \mll\ distribution,
with the templates for prompt leptons obtained from simulation, and those from
misidentified lepton events obtained from
control samples with modified selection cuts. The isolation efficiencies
were measured as a function of lepton \pt\ for four bins in $|\eta|$.
The data results are compared to the corresponding isolation efficiencies
predicted by the {\scr Powheg\,+\,Pythia8} and {\scr Sherpa} \zll\
simulation samples in Figure~\ref{f:leptisol}(a, b). The data efficiencies
in the lowest bin used in the analysis ($25\,\GeV<\pt<30$\,\GeV)
are around 80--85\%,
increasing to 99\% at high lepton \pt. The {\scr Powheg\,+\,Pythia8} simulation
underestimates the electron efficiency by around 2\% at low \pt,
whereas the muon efficiency is overestimated by 1\%. The {\scr Sherpa}
simulation also underestimates the electron efficiency by about 2\% at low \pt,
but overestimates the muon efficiency by about 3\%.
 
\begin{figure}[tp]
\parbox{83mm}{\includegraphics[width=74mm]{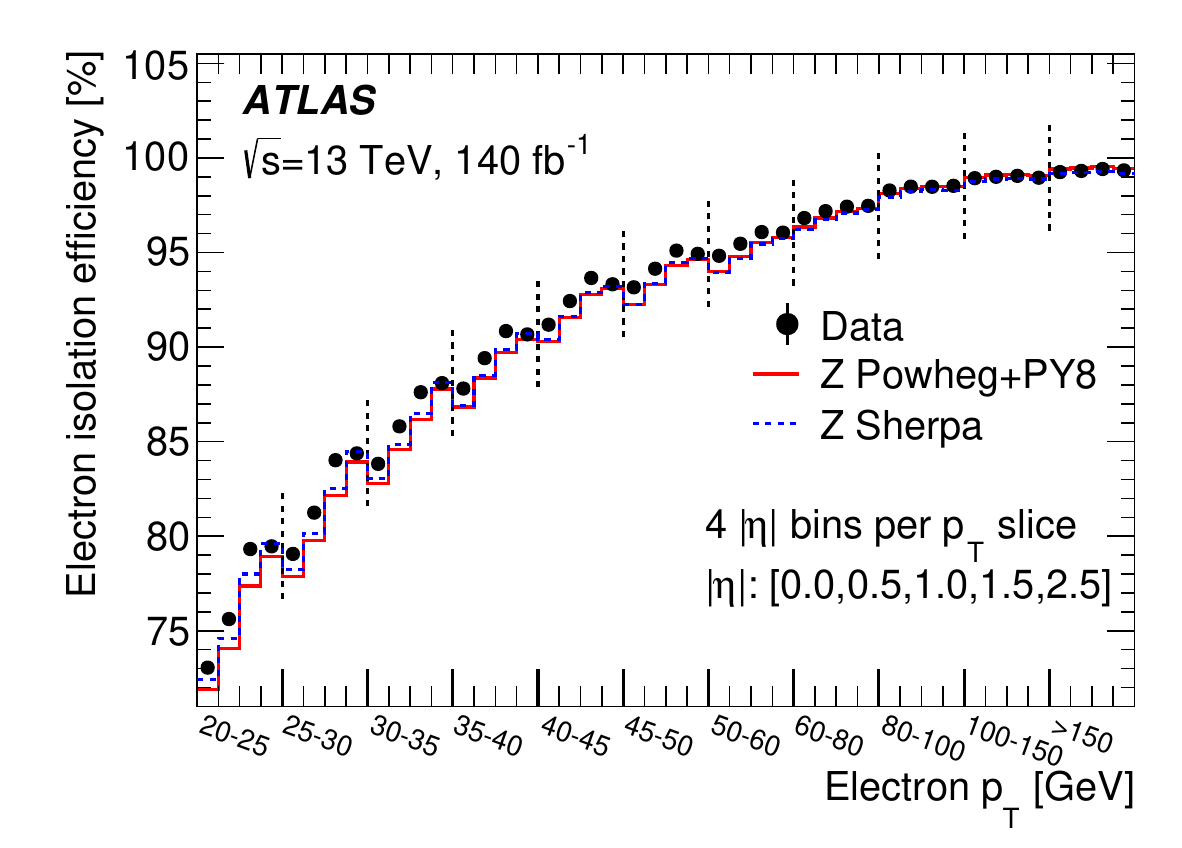}\vspace{-7mm}\center{(a)}}
\parbox{83mm}{\includegraphics[width=74mm]{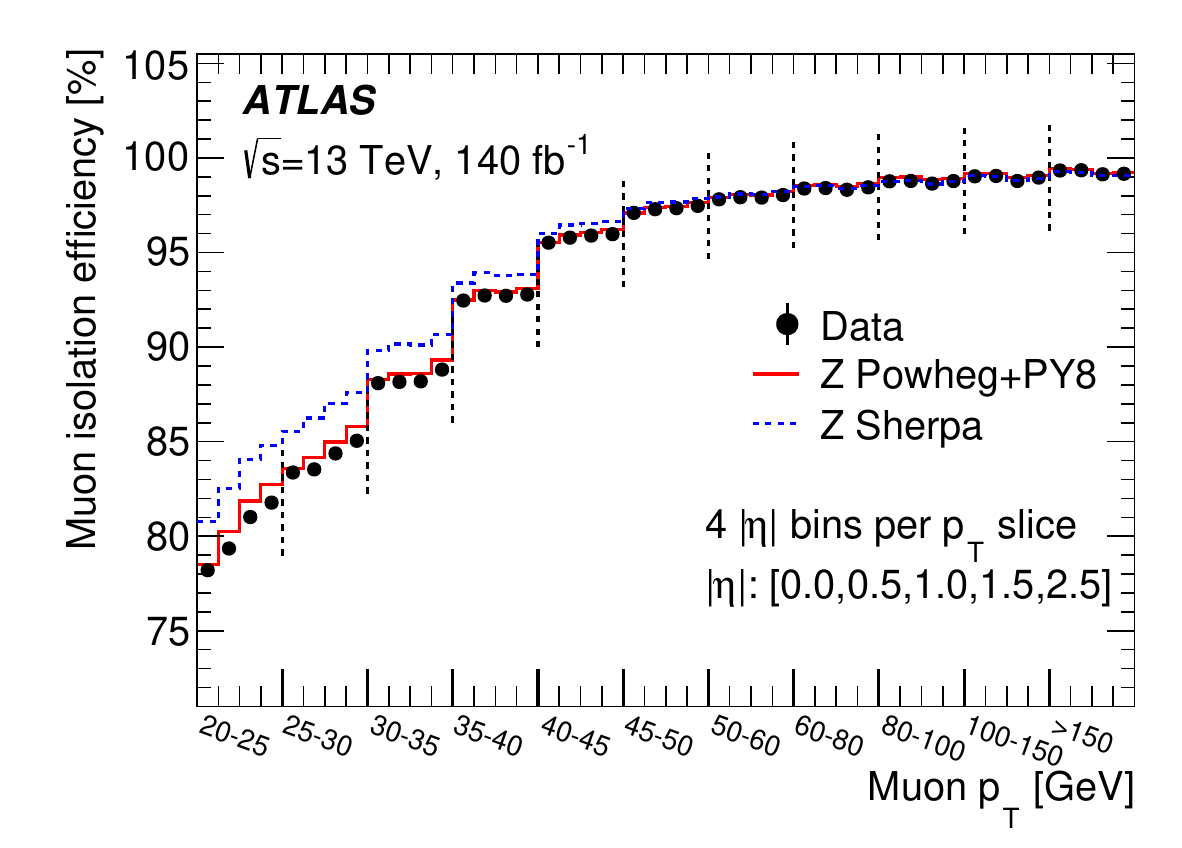}\vspace{-7mm}\center{(b)}}
\parbox{83mm}{\includegraphics[width=74mm]{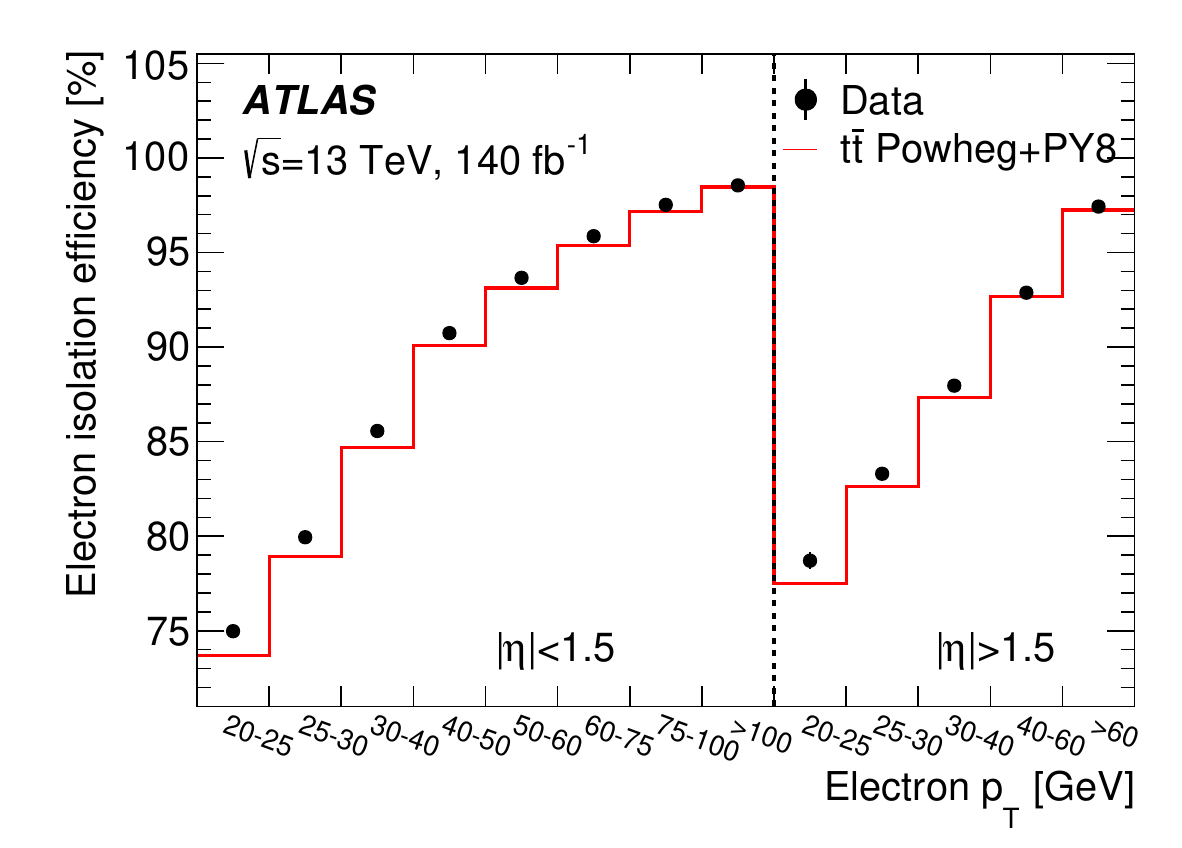}\vspace{-7mm}\center{(c)}}
\parbox{83mm}{\includegraphics[width=74mm]{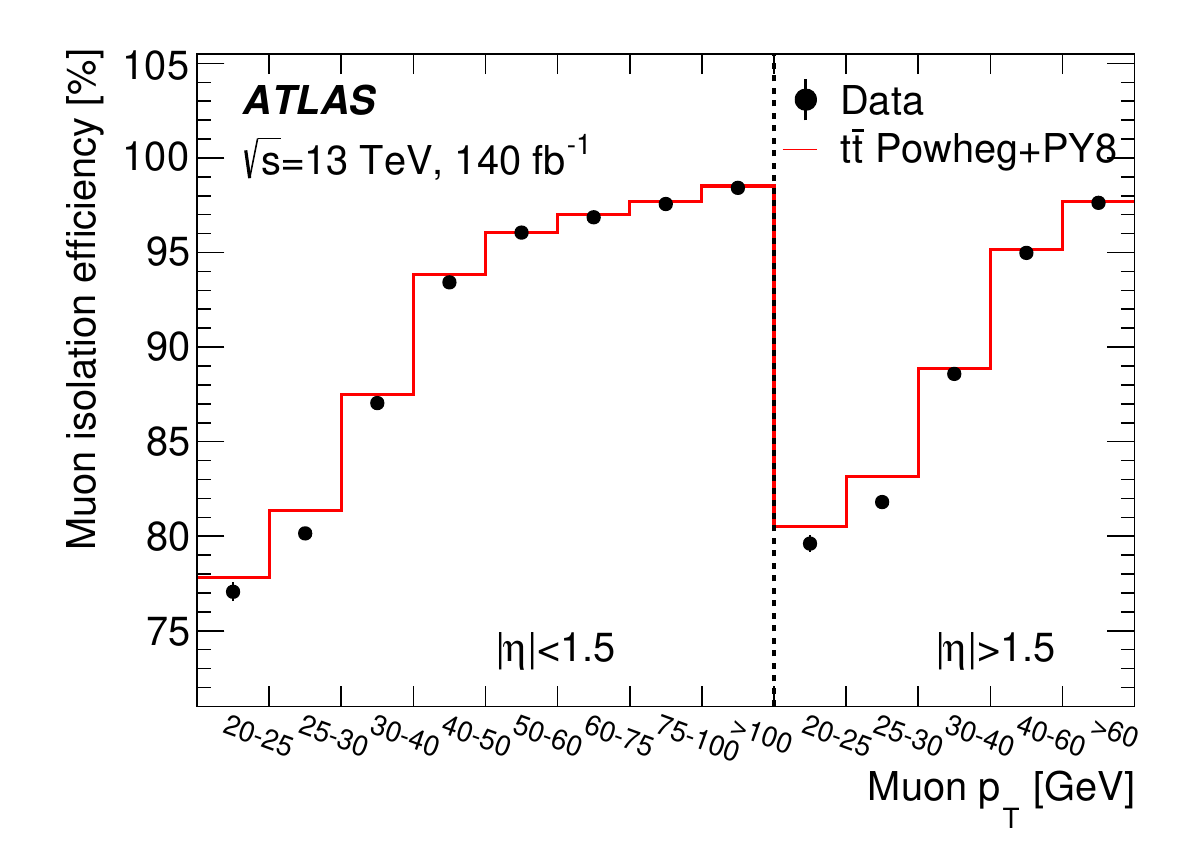}\vspace{-7mm}\center{(d)}}
\caption{\label{f:leptisol}Lepton isolation efficiencies measured for (a, c)
electrons and (b, d) muons as functions of lepton \pt\ and $|\eta|$. The
data is shown by the points with error bars and the simulation by the
histograms. The upper plots show measurements in \zll\ events, with
four $\eta$ bins (with boundaries at $|\eta|=0$, 0.5, 1.0, 1.5 and 2.5)
shown sequentially for each \pt\ slice. The lower
plots show measurements in $\ttbar\rightarrow e\mu$ events, separately for
leptons with $|\eta|<1.5$ in the leftmost bins and $|\eta|>1.5$ in the
rightmost bins. The data is compared to {\scr Powheg\,+\,Pythia8} simulation
(without isolation efficiency scale factors)
for both $Z$ and \ttbar\ events, and additionally to {\scr Sherpa} $Z$ events.}
\end{figure}
 
The corresponding efficiencies in the \ttbar\ environment were measured
using $e\mu$ events with at least one $b$-tagged jet. Here, an invariant
mass cut cannot be applied to suppress misidentified lepton background
and the fraction of background in the probe lepton samples failing isolation
reaches 25--30\% in the samples with one $b$-tagged jet, but
remains below 10\% when two $b$-tagged jets are present. This background
was estimated using same-charge events in a similar way as for the main
analysis selection discussed in Section~\ref{s:meth}. The efficiencies
were measured as a function of lepton \pt\ for $|\eta|<1.5$
and $|\eta|>1.5$, combining the results from the one $b$-tagged and two
$b$-tagged jet samples, which were found to be consistent. The results
from data are compared with the baseline {\scr Powheg\,+\,Pythia8} \ttbar\
simulation in Figure~\ref{f:leptisol}(c, d). In data, the efficiencies
vary from around 80\% at $\pt=25$\,\GeV\ to 97--98\% at high \pt, and the
simulation underestimates the efficiency at low \pt\
for electrons, and overestimates it for muons, similar to the \zll\ case.
 
The ratios of data to simulation efficiencies shown in Figure~\ref{f:leptisol}
were used to define multiplicative efficiency corrections (scale factors)
that were applied to simulation on a
per-lepton basis. The systematic uncertainties on the \zll\ scale factors
were assessed by varying selection cuts for the fit and control regions,
leading to uncertainties of less than 0.05\% on the values of \epszee\
and \epszmm. The systematic uncertainties on the \ttbar\ scale factors
include the modelling of prompt-lepton contributions to the same-charge
samples, the extrapolation of misidentified leptons from same- to
opposite-charge and the modelling of the electron charge misassignment,
giving uncertainties of around 0.1\% per lepton. The isolation efficiency
scale factors calculated for {\scr Powheg\,+\,\Pythia8} \ttbar\ events
were applied to all simulated events passing the \ttbar\ selections,
including the $Z$+jets events simulated with {\scr Sherpa}. However,
the results from inclusive \zmm\ events shown in Figure~\ref{f:leptisol}(b)
suggest that the modelling of the muon isolation efficiency in {\scr Sherpa}
$Z$+jets events may require different corrections to those measured
for {\scr Powheg\,+\Pythia8} \ttbar\ events. A potential extra difference
in lepton isolation efficiencies between $Z(\rightarrow ee)$+jets
and $Z(\rightarrow\mu\mu)$+jets was therefore included in the likelihood
fit, represented by the floating parameter \rzjme\ appearing via
\dzjme\ in Eq.~(\ref{e:zgll}).

\section{Systematic uncertainties}\label{s:syst}
 
Systematic uncertainties arise from uncertainties in the quantities appearing
in Eqs.~(\ref{e:tags}), (\ref{e:tagsf}) and~(\ref{e:ztags}). Each systematic
uncertainty was evaluated by calculating the effect on all input parameters
(\epsllp, \cbllp, \epszll, background estimates etc.)
simultaneously and repeating the fit, rather than including the systematic
uncertainty as a nuisance parameter in the fit in a profile likelihood
approach. For some sources of uncertainty,
e.g.\ lepton efficiencies and PDF variations, the effects on
the different input quantities are only partially correlated, and this was
taken into account by using the full covariance matrix expressing the dependence
of the fit result on each parameter and their correlations. The resulting
systematic uncertainties on \xtt, \xz, \rwzme\ and \rzme\ are discussed below,
and summarised in Table~\ref{t:systsum} in Section~\ref{s:res}.
The methodologies generally follow those described in Ref.~\cite{TOPQ-2018-17}.
 
\begin{description}
\item[$\boldmath{t\bar{t}}$ modelling:] The values of \epsllp, \cbllp,
\ftau{\ell\ell'}{n}\ and
$f_{j,m}^{\ell\ell,t\bar{t}}$ depend on the choice of \ttbar\ simulation model.
The uncertainty due to the choice of generator (in particular the
matrix-element matching algorithm) was assessed by using the sample
generated with {\scr aMC@NLO\,+\,Pythia8} instead of {\scr Powheg\,+\,Pythia8},
and the parton shower, hadronisation and underlying event modelling
uncertainty was assessed by using the {\scr Powheg\,+\,Herwig7.1} sample.
Uncertainties due to initial- and final-state radiation were assessed by
using event weights to
vary the QCD renormalisation and factorisation scales in the matrix
element independently by factors of two up and down from their default values,
by changing the \hdamp\ parameter
from $1.5\mtop$ to $3\mtop$
and symmetrising the resulting uncertainty \cite{ATL-PHYS-PUB-2020-023},
by using the {\scr Var3c} A14 tune variations~\cite{ATL-PHYS-PUB-2014-021},
and by changing the renormalisation
and factorisation scales used in the parton shower, again by factors of
two up and down. Since some of these variations
also induce changes in the amount of activity close to the leptons, they
affect the simulated lepton isolation efficiencies. These variations
were therefore evaluated without applying lepton isolation cuts, to avoid
double-counting differences absorbed in the lepton isolation efficiency
scale factors described in Section~\ref{s:isol}. The fraction of
\ttbar\ events with at least three $b$-jets at generator level was also
varied by $\pm 50$\%, motivated by the discrepancies seen for $\nbtag\geq 3$
in Figure~\ref{f:dmcsf}(a, b),
and the top quark mass was varied by $\pm 1$\,\GeV.
 
\item[Top quark \pt\ modelling:] As none of the considered \ttbar\ modelling
variations reproduce the data lepton \pt\ distribution, the full effect of the
top quark \pt\ reweighting shown as the red dotted line in
Figure~\ref{f:dmcsf}(c,~d) was included as a systematic uncertainty.
 
\item[Parton distribution functions:] The PDF uncertainties were evaluated
using the 30~eigenvectors of the PDF4LHC15 meta-PDF set~\cite{pdf4lhc2},
taking into account the differing effects on the \ttbar, $Wt$, $Z$+jets
and inclusive $Z$ processes and their correlations. The uncertainties
from the PDF4LHC15 eigenvectors are larger than the shifts induced by
reweighting the simulation samples (generated with NNPDF3.0 or CT10 PDFs) to the
central predictions of PDF4LHC15.
 
\item[Single top modelling:] The uncertainties on modelling the $Wt$ background
were assessed using alternative samples generated with
{\scr aMC@NLO\,+\,Pythia8} and {\scr Powheg\,+\,Herwig7.0}~\cite{herwig7},
by varying
the renormalisation and factorisation scales and by using the {\scr Var3c}
tune variations, in the same way as for \ttbar\ events. The $Wt$ cross-section
was varied by the QCD scale uncertainty of 2.4\%, the PDF-related cross-section
uncertainty
being already accounted for coherently with the \ttbar\ and $Z$ PDF variations.
 
\item[Single-top/$\boldmath{t\bar{t}}$ interference:] The uncertainty
in modelling the interference between the $Wt$ and \ttbar\ final states
was assessed by using the diagram subtraction scheme~\cite{wtinter1,wtinter2}
instead of the baseline diagram removal scheme.
 
\item[$Z$(+jets) modelling:] Uncertainties in the $Z$+jets background in
the \ttbar\ selection were evaluated by changing the QCD factorisation and
renormalisation scales in the {\scr Sherpa} samples by factors of two up
and down from their default values, separately or together and
excluding variations in opposite directions. The {\scr Sherpa} samples
were also used to evaluate corresponding variations in \epszee\
and \epszmm\ for the inclusive \zll\ selection.
Simultaneous changes of both scales were found to
have the largest effects on both \rwzme\ and \rzme, and were used to define the
corresponding uncertainty. Half the effect of the \ptll\ weighting applied
to the {\scr Powheg\,+\,Pythia8} inclusive \zll\ samples was included
as an additional $Z$ modelling uncertainty.
 
\item[Diboson modelling:] The normalisation of the diboson contribution was
varied by 20\%, covering uncertainties in the cross-sections and acceptances.
Alternative samples generated using {\scr Powheg\,+\,Pythia8} instead of
{\scr Sherpa} were also considered.
 
\item[Lepton energy/momentum scale and resolution:] The electron energy
scale and muon momentum scale and corresponding resolution were determined
using \zee\ and \zmm\ decays as discussed in
Refs.~\cite{EGAM-2018-01,MUON-2022-01}, and were varied within the
corresponding uncertainties.
 
\item[Lepton identification:] The lepton identification efficiencies were
measured using tag-and-probe techniques applied to \zee\ and \zmm\
events~\cite{EGAM-2021-01, MUON-2018-03}, as functions of lepton \pt\ and $\eta$
for electrons, and $\eta$ and $\phi$ for muons.  The corresponding uncertainties
are only partially correlated across \pt, $\eta$ and $\phi$, and the information
was propagated to \epsllp\ and \epszll\ (and their correlations)
by generating multiple sets of scale factor replicas whose
variations represent the full uncertainty model. The uncertainties
due to electron charge misidentification were studied using \zee\
events and taken into account using the same technique.
 
\item[Lepton isolation:] The lepton isolation efficiency uncertainties
discussed in Section~\ref{s:isol} were also propagated using scale factor
replicas, and taken to be uncorrelated between electrons and muons,
and between the \ttbar\ and \zll\ selections.
 
\item[Lepton trigger:] The lepton trigger efficiencies were also measured
in \zee\ and \zmm\ events using tag-and-probe
techniques~\cite{TRIG-2018-05,TRIG-2018-01}, and were varied within the
corresponding uncertainties.
 
\item[Jet energy scale and resolution:] The jet energy scale was determined
using a combination of simulation, test beam and in-situ measurements,
and the jet energy resolution was studied using di-jet balance
techniques~\cite{JETM-2018-05}. The modelling of the pileup jet veto using the
JVT requirement was studied using jets in \zmm\ events~\cite{PERF-2014-03}.
 
\item[$\boldmath{b}$-tagging efficiency/mistag:] The efficiency for
reconstructing and tagging $b$-jets in \ttbar\ events was measured in situ
via the fit parameters \epsbllp. However, the background yields and tagging
correlations depend on the $b$-tagging efficiencies and mistag rates
predicted by simulation, with the corresponding scale factors and
uncertainties determined using \ttbar\ and $Z$+jets events as described
in Refs.~\cite{FTAG-2018-01,FTAG-2020-08,FTAG-2019-02}.
 
\item[Misidentified leptons:] The uncertainties on the misidentified lepton
backgrounds were evaluated as discussed in Section~\ref{s:meth}, and
were taken to be uncorrelated between electrons and muons, and between
the \ttbar\ and \zll\ selections.
 
\item[Simulation statistics:] The limited size of the Monte Carlo simulation
samples primarily affects the predictions of \epsllp\ and \cbllp, and
the fractions $f_{j,m}^{\ell\ell,Z+\mathrm{jets}}$ of the $Z$+jets background
entering each invariant mass bin. The uncertainties in \epsllp\ and \cbllp\ were
taken into account by repeating the fit with each parameter shifted by
its uncertainty in turn. The uncertainties in $f_{j,m}^{\ell\ell,\mathrm{Z+jets}}$
were accounted for directly in the Gaussian likelihood terms that compare
the data counts with the prediction in each invariant mass  bin.
 
\item[Integrated luminosity:] The integrated luminosity of the dataset
was evaluated using the LUCID2 detector~\cite{LUCID2},
complemented by measurements from the inner detector
and calorimeters, and has an uncertainty of 0.83\%~\cite{DAPR-2021-01}.
 
\item[Beam energy:] The LHC beam energy is known to a precision of 0.1\%,
which translates into small uncertainties on \xtt\ and \xz\ as discussed
in Ref.~\cite{TOPQ-2018-17}.
 
\end{description}
 
\section{Fit results}\label{s:res}
 
Table~\ref{t:fitcounttt} shows the number of observed events in each of the
dilepton channels of the \ttbar\ selection together with the results of the
fit, broken down into the estimated contributions from \ttbar, $Wt$, $Z$+jets,
diboson and misidentified leptons. The counts are shown separately
for events with one and two $b$-tagged jets, and separately for events
off (with $|\mll-m_Z|>10$\,\GeV) and on (with $|\mll-m_Z|<10$\,\GeV)
the $Z$ resonance in the same-flavour channels. The full \mll\ distributions
are shown in Figure~\ref{f:res}. In the same-flavour samples with
one $b$-tagged jet, the \ttbar\ purity is about 70\% and the background
is dominated by $Z$+jets events. In the two $b$-tagged jets samples,
the \ttbar\ purity rises to 93\%, with equal background contributions
from $Z$+jets and $Wt$ events. In the $e\mu$ channel, the $Z$+jets background
is almost negligible, and the \ttbar\ purity is 89\% in the one $b$-tagged
sample and 96\% in the two $b$-tagged sample. Table~\ref{t:fitcountz}
shows the corresponding event counts and predictions for the \zll\
selection; here the backgrounds are only around 0.5\%.
 
\begin{table}
\caption{\label{t:fitcounttt}Observed numbers of opposite-charge
dilepton events (weighted events for the $e\mu$ and $\mu\mu$ channels)
with one (upper block) and two (lower block) $b$-tagged jets
in the \ttbar\ selection in data,
together with the estimated event counts from the fit prediction,
including the associated statistical and systematic uncertainties.
The five columns show the $ee$ channel
with $|\mll-m_Z|>10$\,\GeV\ (off-$Z$) and $|\mll-m_Z|<10$\,\GeV\ (on-$Z$),
the $e\mu$ channel, and the $\mu\mu$ channel including off-$Z$ and on-$Z$
selections. The uncertainties in the total predictions are
smaller than the individual component uncertainties due to
correlations induced by the fit.}
\centering
 
\begin{tabular}{l|
r@{$\,\pm\,$}l
r@{$\,\pm\,$}l|
r@{$\,\pm\,$}l|
r@{$\,\pm\,$}l
r@{$\,\pm\,$}l
}
\hline
Event counts $\rule[-3mm]{0mm}{7mm}$ & \multicolumn{2}{c}{$N_{1,\mathrm{off-Z}}^{ee}$} & \multicolumn{2}{c|}{$N_{1,\mathrm{on-Z}}^{ee}$} & \multicolumn{2}{c|}{$N_{1}^{e\mu}$} & \multicolumn{2}{c}{$N_{1,\mathrm{off-Z}}^{\mu\mu}$} & \multicolumn{2}{c}{$N_{1,\mathrm{on-Z}}^{\mu\mu}$} \\
\hline
Data& \multicolumn{2}{l}{222304}& \multicolumn{2}{l|}{442108}& \multicolumn{2}{l|}{405437}& \multicolumn{2}{l}{223085}& \multicolumn{2}{l}{448105} \\
\hline
\ttbar & 154800 & 1700 & 24830 & 850 & 361000 & 4200 & 152500 & 1800 & 24070 & 860 \\
$Wt$ & 17500 & 1600 & 2770 & 240 & 41500 & 3800 & 17800 & 1700 & 2730 & 250 \\
$Z$+jets & 46880 & 400 & 410700 & 2000 & 859 & 21 & 51010 & 780 & 418000 & 2000 \\
Diboson & 770 & 160 & 3940 & 840 & 790 & 280 & 770 & 160 & 3880 & 830 \\
Mis-ID leptons & 1300 & 500 & 360 & 260 & 1740 & 610 & 390 & 150 & 172 & 87 \\
\hline
Total prediction & 221280 & 550 & 442600 & 1100 & 405900 & 1800 & 222390 & 670 & 448900 & 1100 \\
\hline
\hline
Event counts $\rule[-3mm]{0mm}{7mm}$ & \multicolumn{2}{c}{$N_{2,\mathrm{off-Z}}^{ee}$} & \multicolumn{2}{c|}{$N_{2,\mathrm{on-Z}}^{ee}$} & \multicolumn{2}{c|}{$N_{2}^{e\mu}$} & \multicolumn{2}{c}{$N_{2,\mathrm{off-Z}}^{\mu\mu}$} & \multicolumn{2}{c}{$N_{2,\mathrm{on-Z}}^{\mu\mu}$} \\
\hline
Data& \multicolumn{2}{l}{\phantom{0}85936}& \multicolumn{2}{l|}{\phantom{0}37704}& \multicolumn{2}{l|}{198502}& \multicolumn{2}{l}{\phantom{0}86169}& \multicolumn{2}{l}{\phantom{0}38512} \\
\hline
\ttbar & 79750 & 920 & 13340 & 480 & 191000 & 1800 & 79770 & 830 & 13180 & 450 \\
$Wt$ & 2860 & 760 & 400 & 110 & 6700 & 1600 & 2940 & 740 & 423 & 90 \\
$Z$+jets & 2675 & 68 & 23610 & 590 & 78 & 2 & 3095 & 87 & 24110 & 600 \\
Diboson & 67 & 23 & 550 & 110 & 29 & 8 & 71 & 30 & 570 & 110 \\
Mis-ID leptons & 400 & 290 & 96 & 59 & 720 & 520 & 350 & 160 & 104 & 56 \\
\hline
Total prediction & 85760 & 360 & 38000 & 190 & 198510 & 440 & 86230 & 300 & 38380 & 210 \\
\hline
\end{tabular}
\end{table}
 
\begin{table}
\caption{\label{t:fitcountz}Observed numbers of opposite-charge dilepton
events (weighted events for the $\mu\mu$ channel)
in the inclusive \zee\ and \zmm\ selections in data, together with the
estimated event counts from the fit prediction,
including the associated
statistical and systematic uncertainties. The uncertainties in the total
predictions are much smaller than the individual uncertainties due to
correlations induced by the fit.}
\centering
 
\begin{tabular}{l|
r@{$\,\pm\,$}l
r@{$\,\pm\,$}l
}\hline
Event counts & \multicolumn{2}{c}{\zee} & \multicolumn{2}{c}{\zmm} \\
\hline
Data & \multicolumn{2}{l}{47898836} & \multicolumn{2}{l}{49016812} \\
\hline
\zll & 47621000 & 33000 & 48767000 & 29000 \\
Diboson & 111000 & 22000 & 104000 & 21000 \\
\ztt & 16850 & 140 & 13780 & 110 \\
\ttbar & 119000 & 14000 & 117000 & 14000 \\
$Wt$ & 12380 & 890 & 12390 & 880 \\
Mis-ID leptons & 19000 & 18000 & 3000 & 13000 \\
\hline
Total prediction & 47898800 & 6900 & 49016800 & 6200\\
\hline
\end{tabular}
\end{table}
 
Figure~\ref{f:res} shows that the fit generally models the data well, except
for a data excess in the lowest \mll\ bin in all four same-flavour
distributions, whose significance is however always less than two standard
deviations. Part of this discrepancy can be attributed to the top-quark \pt\
mismodelling---the reweighted sample shown by the red dotted line in
Figure~\ref{f:dmcsf} also predicts a softer \mll\ distribution for
\ttbar\ events, as also observed for the \mll\ distribution in the
\ttbar-dominated $e\mu$ channel. The effect of this reweighting
defines the `Top quark \pt\ modelling' uncertainty in
Table~\ref{t:systsum} and is included in the cyan bands shown in
Figure~\ref{f:res}.
 
\begin{figure}[tp]
\parbox{83mm}{\includegraphics[width=74mm]{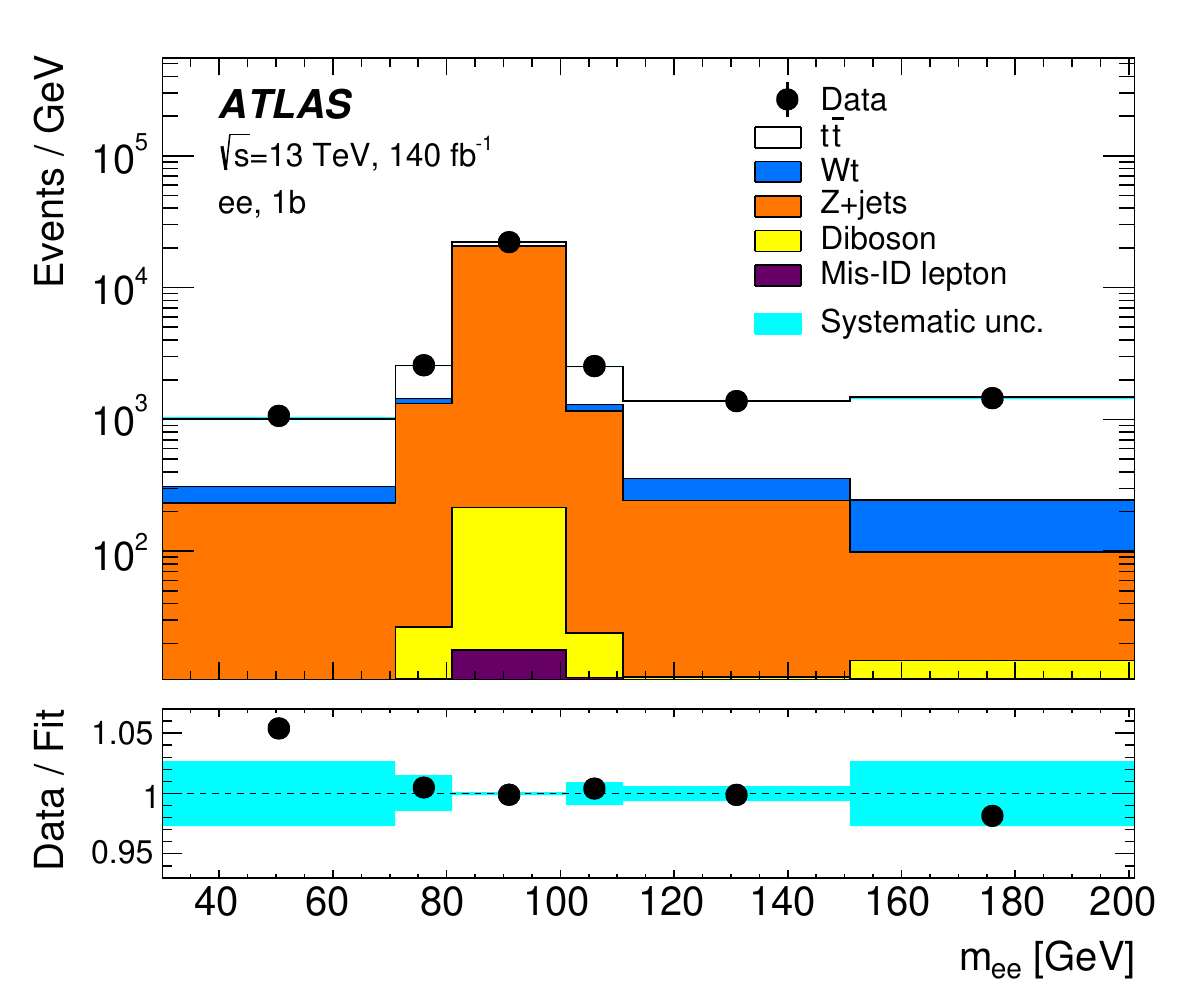}\vspace{-7mm}\center{(a)}}
\parbox{83mm}{\includegraphics[width=74mm]{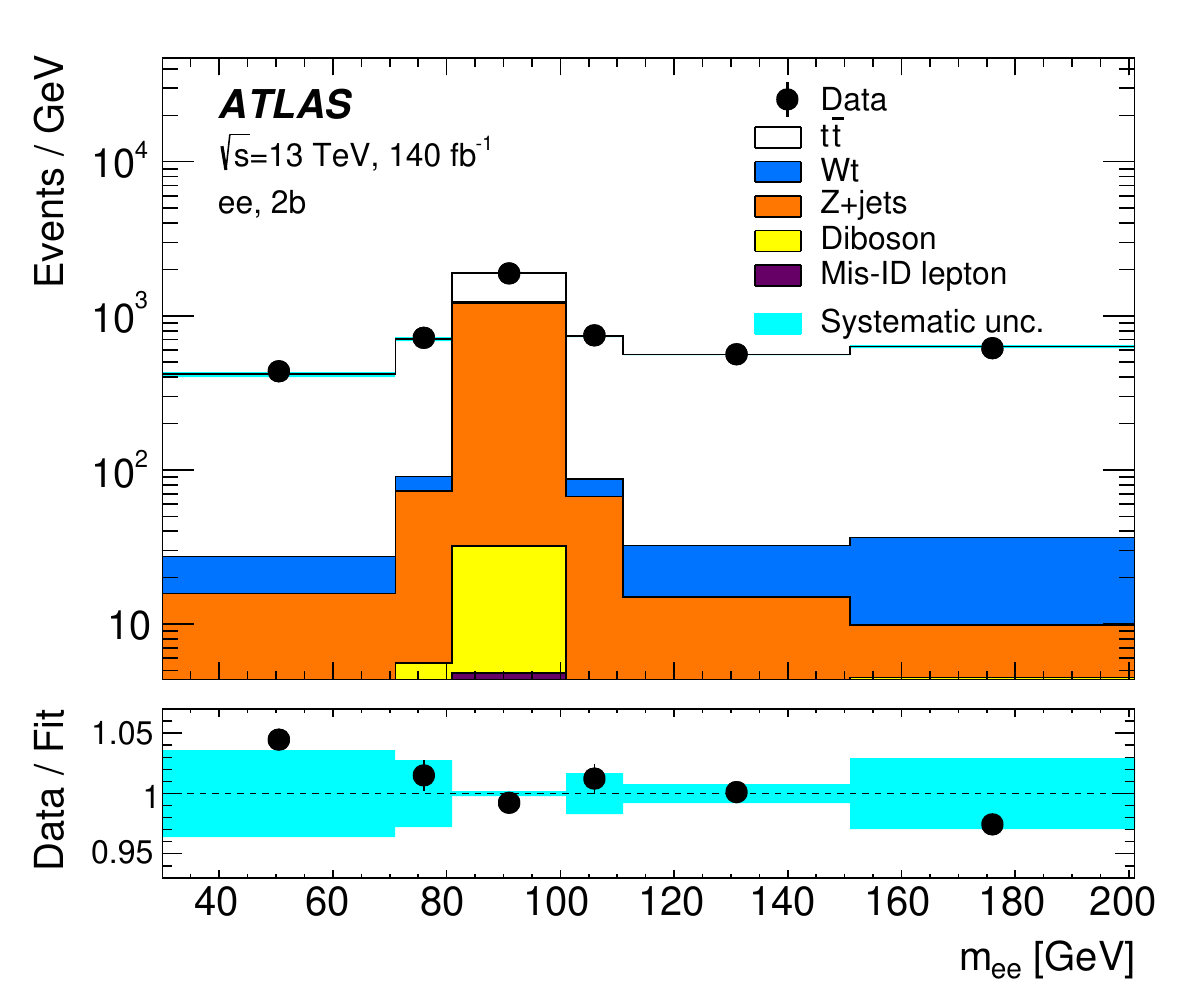}\vspace{-7mm}\center{(b)}}
\parbox{83mm}{\includegraphics[width=74mm]{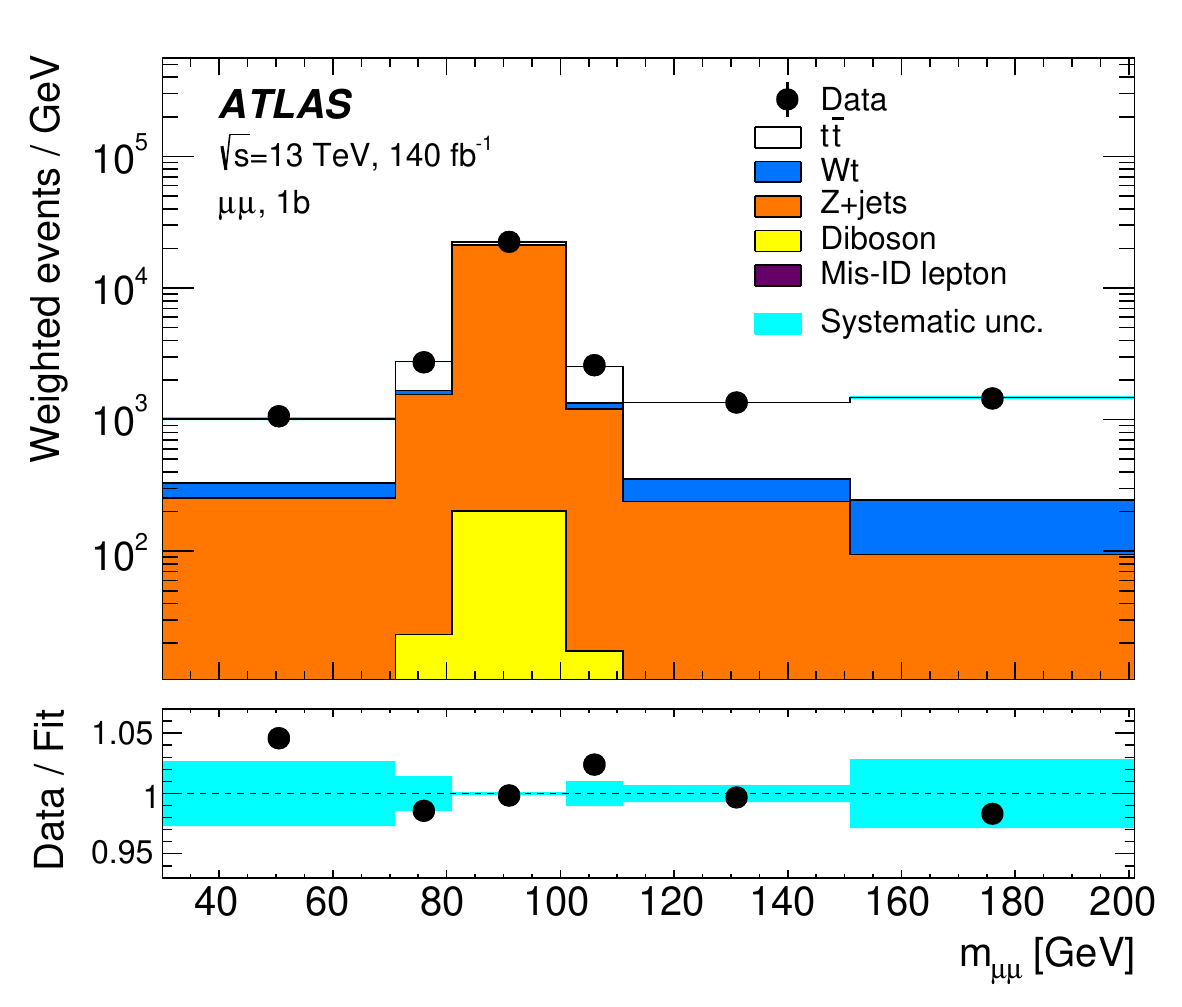}\vspace{-7mm}\center{(c)}}
\parbox{83mm}{\includegraphics[width=74mm]{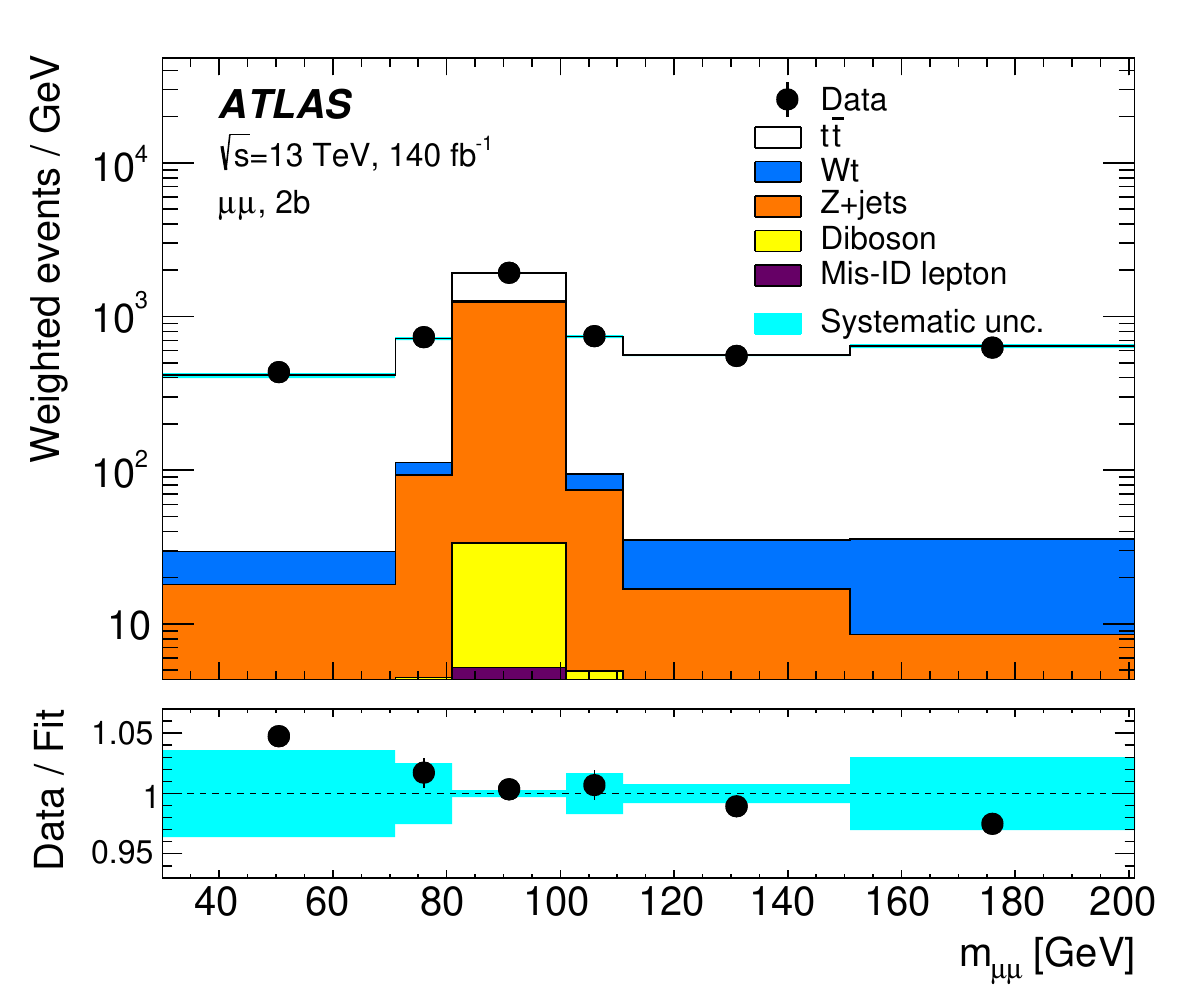}\vspace{-7mm}\center{(d)}}
\parbox{83mm}{\includegraphics[width=74mm]{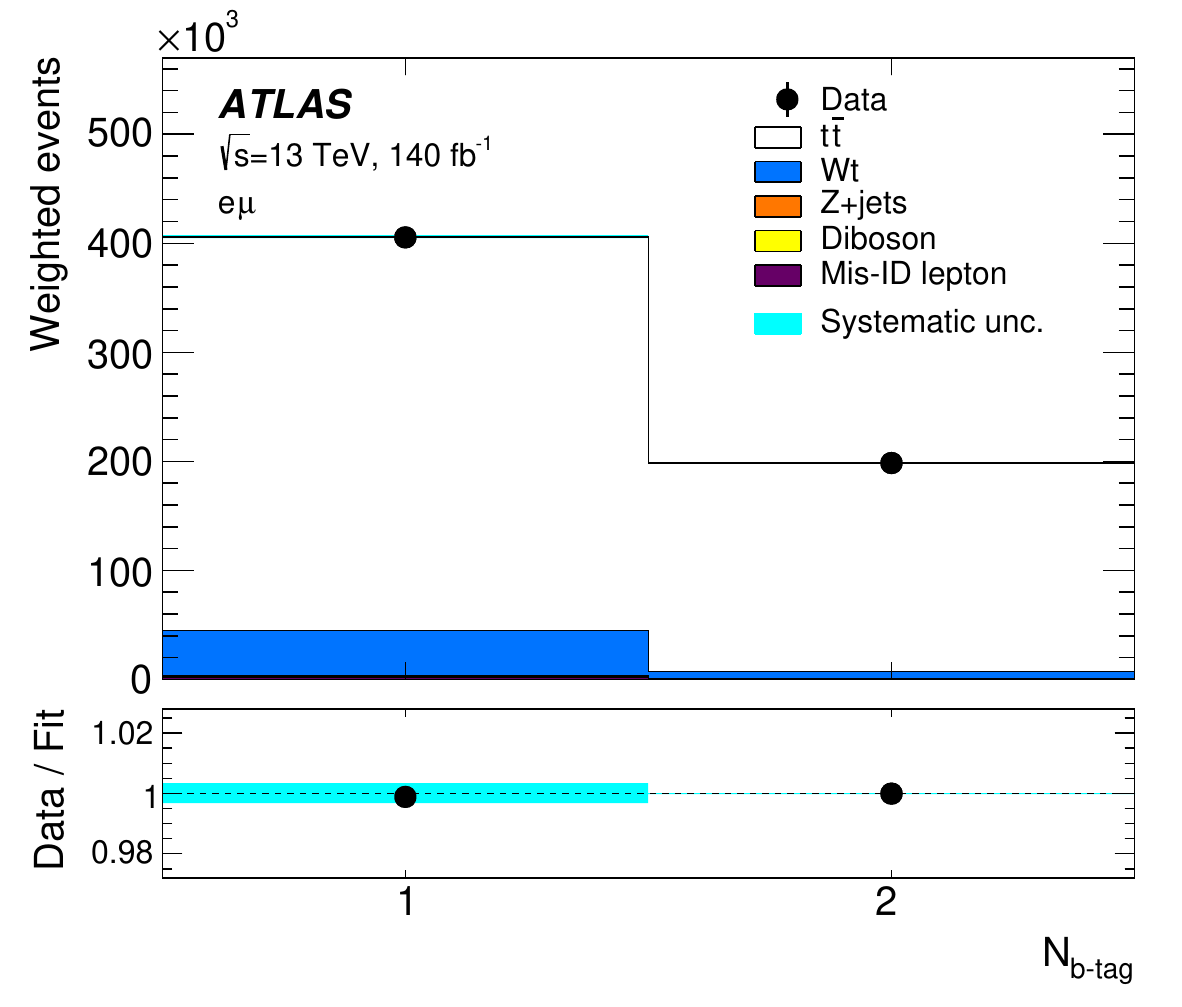}\vspace{-7mm}\center{(e)}}
\parbox{83mm}{\includegraphics[width=74mm]{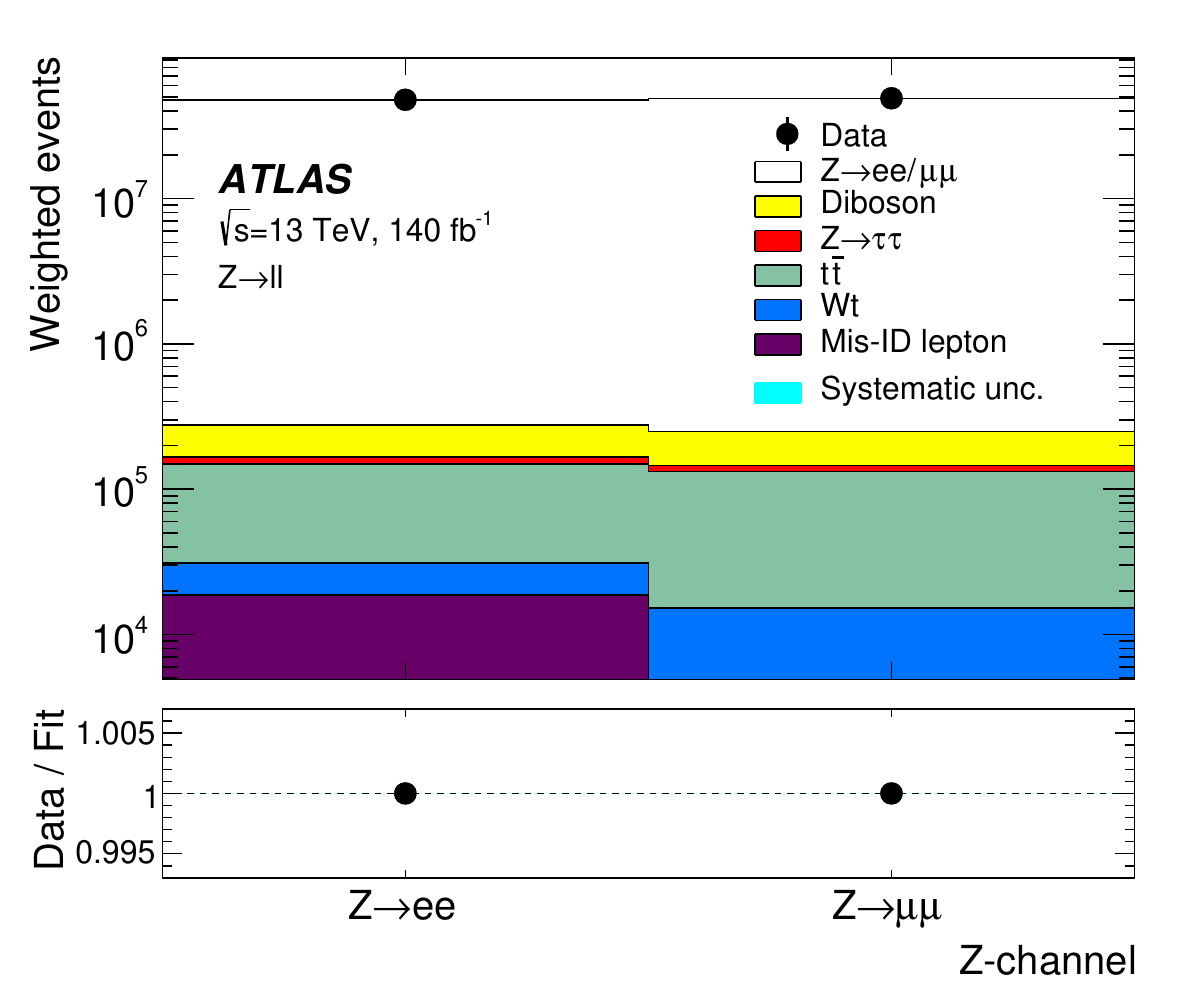}\vspace{-7mm}\center{(f)}}
\caption{\label{f:res}Results of the fit to data, showing the invariant
mass distributions for the one and two $b$-tagged jet samples in the
(a, b) $ee$ and (c, d) $\mu\mu$ channels, (e) the $b$-tagged jet multiplicity
in the $e\mu$ channel, and (f) the number of events in the inclusive \zll\
selection. The data are shown by the points with statistical error bars, and
are compared with the results of the fit, showing the scaled contributions
from \ttbar, $Wt$, $Z$+jets, dibosons, $Z\rightarrow\tau\tau$ and events with
misidentified leptons. The total systematic uncertainty of the fit prediction
in each bin is shown by the cyan band. The lower panels show the ratios of
data to the fit predictions. In the invariant mass distributions,
the last bin includes the overflow with $\mll>200$\,\GeV\ but is normalised
to the displayed bin width.}
\end{figure}
 
The fit results for the cross-sections are
\begin{eqnarray*}
\xtt & = &  \phz 809.5 \pm  1.1 \pm 20.1 \pm  \phz 7.5 \pm  1.9\,\mathrm{pb}\ ,\\
\xz & = &  2019.4 \pm   0.2 \pm  20.7 \pm  16.8 \pm   1.8\,\mathrm{pb}\ ,
\end{eqnarray*}
where the four uncertainties are due to data statistics, systematic effects,
and the knowledge of the integrated luminosity and the LHC beam energy.
As shown in Table~\ref{t:systsum},
the \ttbar\ cross-section result has a precision of \xttrunc\%, dominated by
the uncertainties from \ttbar\ modelling, the top-quark $p_T$ modelling
and the integrated luminosity. It is compatible with the theoretical
prediction discussed in Section~\ref{s:datmc} and with the result
from the $e\mu$ channel alone reported in Ref.~\cite{TOPQ-2018-26},
taking into account the larger systematic uncertainty in this analysis
due to the use of the same-flavour channels and the tighter lepton \pt\
requirement. The result for \xz\ represents the inclusive cross-section
for $Z/\gamma^*\rightarrow\ell\ell$ production for a single
dilepton flavour with $\mll>60$\,\GeV.
In order to compare with previous measurements, it was translated
into a fiducial cross-section \xzfid\ requiring two Born-level leptons with
$\pt>25$\,\GeV and $|\eta|<2.5$, and $66<\mll<116$\,\GeV. The relationship
between the total and fiducial cross-sections is given by
$\xzfid=A_Z\xz$, and the factor $A_Z=0.3836\pm 0.0005$ was evaluated from the
{\scr Powheg\,+\,Pythia8} \zll\ sample, including the extrapolation to the
lower lepton \pt\ requirement of $\pt>25$\,\GeV. The resulting
fiducial cross-section is
\begin{equation}\nonumber
\xzfid =   774.7 \pm   0.1 \pm   1.8 \pm   6.4 \pm   0.7\,\mathrm{pb}\ .
\end{equation}
The systematic uncertainty is much smaller than that for \xz\ because of
strong reductions in the PDF and $Z$ modelling uncertainties in the fiducial
cross-section measurement. The result is compatible with that measured
in Ref.~\cite{STDM-2018-17}.
 
\begin{table}
\caption{\label{t:systsum}Breakdown of the statistical and systematic
uncertainties on the measured cross-sections \xtt\ and \xz, and on the ratios
of branching ratios \rwzme\ and \rzme.}
\centering
 
\begin{tabular}{l|cc|cc}\hline
Uncertainty [\%] & \xtt& \xz& $\rwzme$& $\rzme$\\
\hline
Data statistics& 0.13& 0.01& 0.22& 0.02\\
\hline
\ttbar\ modelling& 1.68& 0.03& 0.10& 0.00\\
Top-quark \pt\ modelling& 1.42& 0.00& 0.06& 0.00\\
Parton distribution functions& 0.67& 0.68& 0.15& 0.03\\
Single-top modelling& 0.65& 0.00& 0.05& 0.00\\
Single-top/\ttbar\ interference& 0.54& 0.00& 0.09& 0.00\\
$Z$(+jets) modelling& 0.06& 0.73& 0.13& 0.20\\
Diboson modelling& 0.05& 0.04& 0.01& 0.00\\
Electron energy scale/resolution& 0.05& 0.06& 0.10& 0.11\\
Electron identification& 0.10& 0.07& 0.04& 0.13\\
Electron charge misidentification& 0.06& 0.06& 0.01& 0.13\\
Electron isolation& 0.09& 0.02& 0.08& 0.04\\
Muon momentum scale/resolution& 0.04& 0.02& 0.06& 0.04\\
Muon identification& 0.18& 0.12& 0.11& 0.23\\
Muon isolation& 0.09& 0.01& 0.07& 0.01\\
Lepton trigger& 0.09& 0.12& 0.01& 0.23\\
Jet energy scale/resolution& 0.08& 0.00& 0.03& 0.00\\
$b$-tagging efficiency/mistag& 0.14& 0.00& 0.00& 0.00\\
Misidentified leptons& 0.17& 0.02& 0.15& 0.05\\
Simulation statistics& 0.04& 0.00& 0.06& 0.00\\
Integrated luminosity& 0.93& 0.83& 0.00& 0.00\\
Beam energy& 0.23& 0.09& 0.00& 0.00\\
\hline
Total uncertainty& 2.66& 1.32& 0.42& 0.45\\
\hline
\end{tabular}
\end{table}
 
The values of \epsbllp\ for the three dilepton channels were found to be
compatible with the values expected from simulation, and are all close to 0.51.
The $Z$+jets scaling parameters were measured to be
$\rza=0.89\pm 0.09$ and $\rzb=1.12\pm 0.32$, the uncertainties being dominated
by the QCD scale variations in the $Z$+jets samples, which significantly change
the predicted cross-sections. The $Z$+jets lepton isolation efficiency
difference was fitted as $\rzjme=0.990\pm 0.003$, compatible with the
differences between {\scr Sherpa} and {\scr Powheg\,+\,Pythia8} lepton
isolation efficiencies shown for inclusive \zll\ events in
Figure~\ref{f:leptisol}.
 
The two ratios of branching ratios were fitted to be
\begin{eqnarray}
\rwzme & = &  0.9990 \pm 0.0022 \pm 0.0036 \nonumber\ , \\
\rzme & = &  0.9913 \pm  0.0002 \pm 0.0045 \nonumber\ ,
\end{eqnarray}
where the first uncertainties are statistical and the second systematic.
A detailed breakdown of the uncertainties is shown in Table~\ref{t:systsum}.
The value of \rzme\ is \rzmesig\ standard deviations below unity, hinting at
a potential bias in the electron or muon identification efficiencies.
The normalisation of \rwzme\ by \rzme\ via Eq.~(\ref{e:rwzme}) protects
\rwzme\ against such a bias, modulo differences in the lepton \pt\ and $\eta$
distributions in dilepton \ttbar\ and \zll\ events.
 
Consistent results were found when analysing the 2015--16, 2017 and~2018
datasets separately.
The result for \rwzme\ was found to be stable when tightening the lepton
\pt\ requirement progressively up to $\pt>40$\,\GeV, and when tightening
the $\eta$ requirement to $|\eta|$<1.5, in each case removing around 40\%
of the \ttbar\ sample. It also changed by less than 0.01\% when removing the
lowest \mll\ bin from the fit, demonstrating insensitivity to the
mismodelling shown in Figure~\ref{f:res}. This mismodelling is consistent
between $ee$ and $\mu\mu$ channels, as can be seen from Figure~\ref{f:resrat},
which shows the ratio of $\mu\mu$ to $ee$ events in each invariant mass bin,
cancelling any common mismodelling. The data and fit predictions for this
ratio agree well in all \mll\ bins.
 
\begin{figure}[tp]
\parbox{83mm}{\includegraphics[width=74mm]{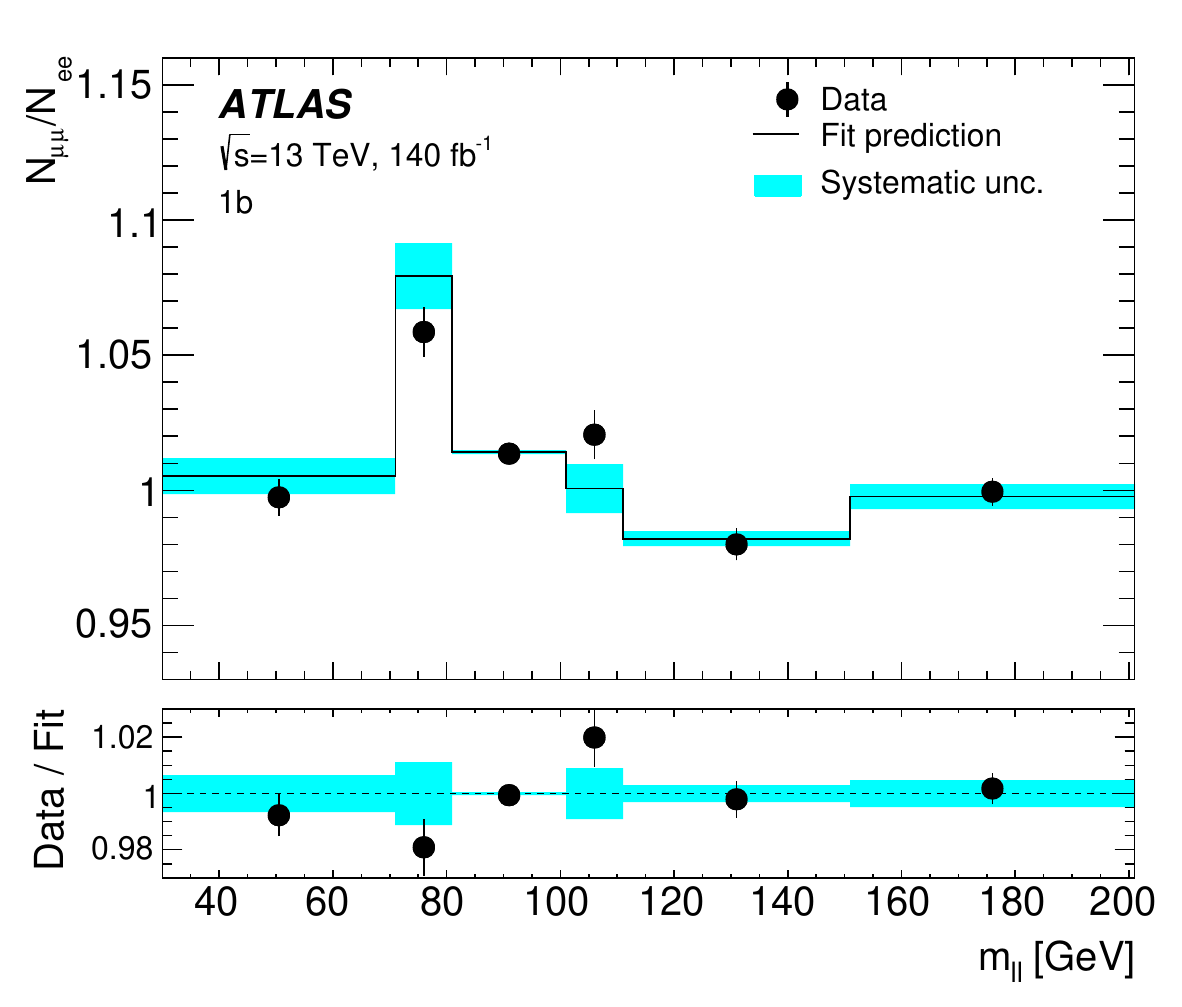}\vspace{-7mm}\center{(a)}}
\parbox{83mm}{\includegraphics[width=74mm]{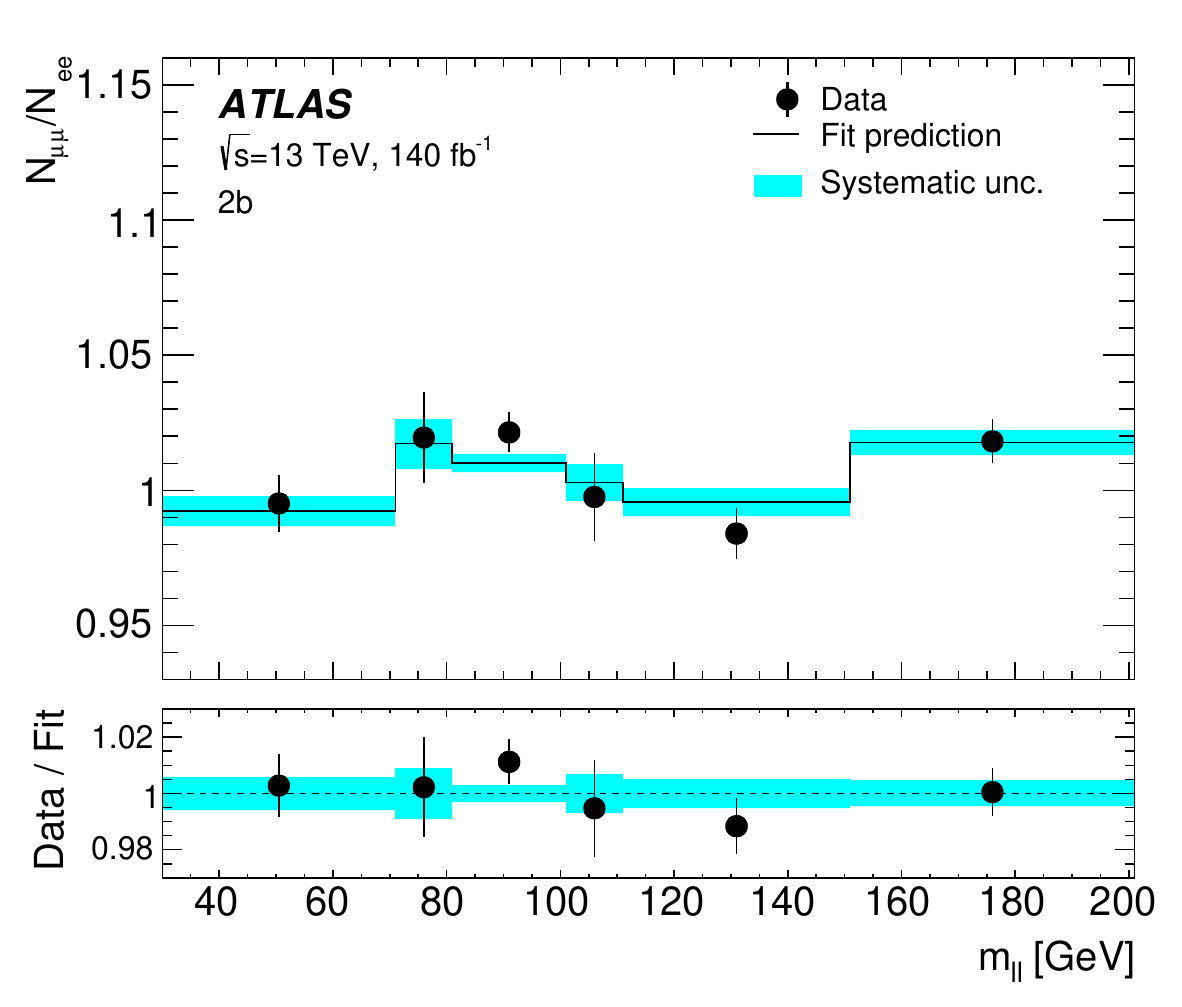}\vspace{-7mm}\center{(b)}}
\caption{\label{f:resrat}Ratio of the number of events in the $\mu\mu$ channel
divided by that in the $ee$ channel as a function of dilepton invariant mass
for events with (a) one and (b) two $b$-tagged jets. The ratio in data
is shown by the points with statistical error bars,
and the results of the fit prediction by the solid lines,
with the cyan band indicating the systematic uncertainty.}
\end{figure}
 
\begin{figure}[tp]
\centering
\includegraphics[width=100mm]{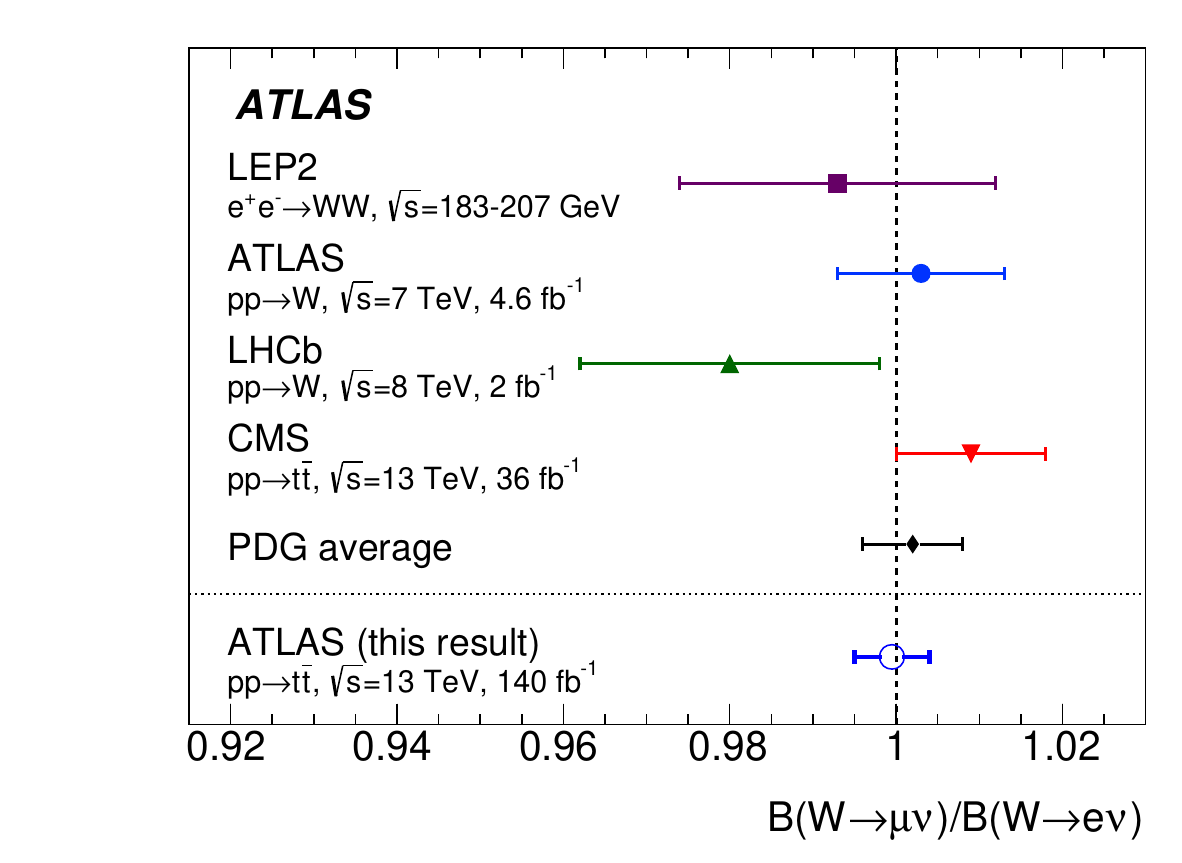}
\caption{\label{f:barplot}Measurement of $\rwme=\brwm/\brwe$ from this analysis compared
to previous results from LEP2 and LHC
experiments~\cite{lepwbr,STDM-2012-20,lhcbwbr,CMS-SMP-18-011} and the
Particle Data Group average~\cite{pdg}.}
\end{figure}
 
The measured value of \rwzme\ was converted to \rwme\ by using the external
measurement of $\rzmeext=1.0009\pm 0.0028$ from LEP and
SLD~\cite{lepzew,pdg}, giving a result of
\begin{equation}\nonumber
\rwme = \rwzme\sqrt{\rzmeext} = \rwmeval \pm \rwmestat\,\mathrm{(stat)}\,\pm \rwmesyst\,\mathrm{(syst)}\,\pm \rwmeext\,\mathrm{(ext)}\ ,
\end{equation}
where the three uncertainties correspond to data statistics, systematic
uncertainties from this analysis, and the uncertainty on the value of \rzmeext
(considered uncorrelated),
giving a total uncertainty of \rwmetot. The result is consistent with the
assumption of lepton flavour universality and with previous measurements,
and has higher precision than the previous world average~\cite{pdg}.
The result is compared with previous measurements of \rwme\ in
Figure~\ref{f:barplot}.
 
\FloatBarrier
\section{Conclusion}\label{s:conc}
 
The ratio of branching ratios $\rwme=\brwm/\brwe$ has been determined using
the complete ATLAS Run~2 \sxyt\ $pp$ collision data sample recorded at the LHC,
by measuring the \ttbar\
cross-section in the $ee$, $e\mu$ and $\mu\mu$ dilepton channels. Systematic
uncertainties due to lepton identification and trigger efficiencies were
minimised by normalising the result to a simultaneous measurement
of $\rzme=\brzmm/\brzee$, and utilising the high-precision measurement
of \rzme\ by the LEP and SLD collaborations. The resulting value of
$\rwme=\rwmeval\pm\rwmetot$ is consistent with the assumption of
lepton flavour universality. This is the most precise measurement
of \rwme\ to date, with a smaller uncertainty than the previous world average.
 
\section*{Acknowledgements}


We thank CERN for the very successful operation of the LHC and its injectors, as well as the support staff at
CERN and at our institutions worldwide without whom ATLAS could not be operated efficiently.
 
The crucial computing support from all WLCG partners is acknowledged gratefully, in particular from CERN, the ATLAS Tier-1 facilities at TRIUMF/SFU (Canada), NDGF (Denmark, Norway, Sweden), CC-IN2P3 (France), KIT/GridKA (Germany), INFN-CNAF (Italy), NL-T1 (Netherlands), PIC (Spain), RAL (UK) and BNL (USA), the Tier-2 facilities worldwide and large non-WLCG resource providers. Major contributors of computing resources are listed in Ref.~\cite{ATL-SOFT-PUB-2023-001}.
 
We gratefully acknowledge the support of ANPCyT, Argentina; YerPhI, Armenia; ARC, Australia; BMWFW and FWF, Austria; ANAS, Azerbaijan; CNPq and FAPESP, Brazil; NSERC, NRC and CFI, Canada; CERN; ANID, Chile; CAS, MOST and NSFC, China; Minciencias, Colombia; MEYS CR, Czech Republic; DNRF and DNSRC, Denmark; IN2P3-CNRS and CEA-DRF/IRFU, France; SRNSFG, Georgia; BMBF, HGF and MPG, Germany; GSRI, Greece; RGC and Hong Kong SAR, China; ISF and Benoziyo Center, Israel; INFN, Italy; MEXT and JSPS, Japan; CNRST, Morocco; NWO, Netherlands; RCN, Norway; MNiSW, Poland; FCT, Portugal; MNE/IFA, Romania; MESTD, Serbia; MSSR, Slovakia; ARIS and MVZI, Slovenia; DSI/NRF, South Africa; MICIU/AEI, Spain; SRC and Wallenberg Foundation, Sweden; SERI, SNSF and Cantons of Bern and Geneva, Switzerland; NSTC, Taipei; TENMAK, T\"urkiye; STFC/UKRI, United Kingdom; DOE and NSF, United States of America.
 
Individual groups and members have received support from BCKDF, CANARIE, CRC and DRAC, Canada; PRIMUS 21/SCI/017, CERN-CZ and FORTE, Czech Republic; COST, ERC, ERDF, Horizon 2020, ICSC-NextGenerationEU and Marie Sk{\l}odowska-Curie Actions, European Union; Investissements d'Avenir Labex, Investissements d'Avenir Idex and ANR, France; DFG and AvH Foundation, Germany; Herakleitos, Thales and Aristeia programmes co-financed by EU-ESF and the Greek NSRF, Greece; BSF-NSF and MINERVA, Israel; Norwegian Financial Mechanism 2014-2021, Norway; NCN and NAWA, Poland; La Caixa Banking Foundation, CERCA Programme Generalitat de Catalunya and PROMETEO and GenT Programmes Generalitat Valenciana, Spain; G\"{o}ran Gustafssons Stiftelse, Sweden; The Royal Society and Leverhulme Trust, United Kingdom.
 
In addition, individual members wish to acknowledge support from CERN: European Organization for Nuclear Research (CERN PJAS); Chile: Agencia Nacional de Investigaci\'on y Desarrollo (FONDECYT 1190886, FONDECYT 1210400, FONDECYT 1230812, FONDECYT 1230987); China: Chinese Ministry of Science and Technology (MOST-2023YFA1605700), National Natural Science Foundation of China (NSFC - 12175119, NSFC 12275265, NSFC-12075060); Czech Republic: Czech Science Foundation (GACR - 24-11373S), Ministry of Education Youth and Sports (FORTE CZ.02.01.01/00/22\_008/0004632), PRIMUS Research Programme (PRIMUS/21/SCI/017); EU: H2020 European Research Council (ERC - 101002463); European Union: European Research Council (ERC - 948254, ERC 101089007), Horizon 2020 Framework Programme (MUCCA - CHIST-ERA-19-XAI-00), European Union, Future Artificial Intelligence Research (FAIR-NextGenerationEU PE00000013), Italian Center for High Performance Computing, Big Data and Quantum Computing (ICSC, NextGenerationEU); France: Agence Nationale de la Recherche (ANR-20-CE31-0013, ANR-21-CE31-0013, ANR-21-CE31-0022, ANR-22-EDIR-0002), Investissements d'Avenir Labex (ANR-11-LABX-0012); Germany: Baden-Württemberg Stiftung (BW Stiftung-Postdoc Eliteprogramme), Deutsche Forschungsgemeinschaft (DFG - 469666862, DFG - CR 312/5-2); Italy: Istituto Nazionale di Fisica Nucleare (ICSC, NextGenerationEU), Ministero dell'Università e della Ricerca (PRIN - 20223N7F8K - PNRR M4.C2.1.1); Japan: Japan Society for the Promotion of Science (JSPS KAKENHI JP21H05085, JSPS KAKENHI JP22H01227, JSPS KAKENHI JP22H04944, JSPS KAKENHI JP22KK0227); Netherlands: Netherlands Organisation for Scientific Research (NWO Veni 2020 - VI.Veni.202.179); Norway: Research Council of Norway (RCN-314472); Poland: Ministry of Science and Higher Education (IDUB AGH, POB8, D4 no 9722), Polish National Agency for Academic Exchange (PPN/PPO/2020/1/00002/U/00001), Polish National Science Centre (NCN 2021/42/E/ST2/00350, NCN OPUS nr 2022/47/B/ST2/03059, NCN UMO-2019/34/E/ST2/00393, UMO-2020/37/B/ST2/01043, UMO-2021/40/C/ST2/00187, UMO-2022/47/O/ST2/00148, UMO-2023/49/B/ST2/04085); Slovenia: Slovenian Research Agency (ARIS grant J1-3010); Spain: Generalitat Valenciana (Artemisa, FEDER, IDIFEDER/2018/048), Ministry of Science and Innovation (MCIN \& NextGenEU PCI2022-135018-2, MICIN \& FEDER PID2021-125273NB, RYC2019-028510-I, RYC2020-030254-I, RYC2021-031273-I, RYC2022-038164-I), PROMETEO and GenT Programmes Generalitat Valenciana (CIDEGENT/2019/023, CIDEGENT/2019/027); Sweden: Swedish Research Council (Swedish Research Council 2023-04654, VR 2018-00482, VR 2022-03845, VR 2022-04683, VR 2023-03403, VR grant 2021-03651), Knut and Alice Wallenberg Foundation (KAW 2018.0157, KAW 2018.0458, KAW 2019.0447, KAW 2022.0358); Switzerland: Swiss National Science Foundation (SNSF - PCEFP2\_194658); United Kingdom: Leverhulme Trust (Leverhulme Trust RPG-2020-004), Royal Society (NIF-R1-231091); United States of America: U.S. Department of Energy (ECA DE-AC02-76SF00515), Neubauer Family Foundation.


\clearpage
 
\printbibliography
 
\clearpage
 
\begin{flushleft}
\hypersetup{urlcolor=black}
{\Large The ATLAS Collaboration}

\bigskip

\AtlasOrcid[0000-0002-6665-4934]{G.~Aad}$^\textrm{\scriptsize 103}$,
\AtlasOrcid[0000-0001-7616-1554]{E.~Aakvaag}$^\textrm{\scriptsize 16}$,
\AtlasOrcid[0000-0002-5888-2734]{B.~Abbott}$^\textrm{\scriptsize 121}$,
\AtlasOrcid[0000-0002-0287-5869]{S.~Abdelhameed}$^\textrm{\scriptsize 117a}$,
\AtlasOrcid[0000-0002-1002-1652]{K.~Abeling}$^\textrm{\scriptsize 55}$,
\AtlasOrcid[0000-0001-5763-2760]{N.J.~Abicht}$^\textrm{\scriptsize 49}$,
\AtlasOrcid[0000-0002-8496-9294]{S.H.~Abidi}$^\textrm{\scriptsize 29}$,
\AtlasOrcid[0009-0003-6578-220X]{M.~Aboelela}$^\textrm{\scriptsize 44}$,
\AtlasOrcid[0000-0002-9987-2292]{A.~Aboulhorma}$^\textrm{\scriptsize 35e}$,
\AtlasOrcid[0000-0001-5329-6640]{H.~Abramowicz}$^\textrm{\scriptsize 152}$
% ...
\bigskip
\\

% ...
$^{16}$Department for Physics and Technology, University of Bergen, Bergen; Norway.\\
% ...
$^{29}$Physics Department, Brookhaven National Laboratory, Upton NY; United States of America.\\
% ...
$^{35}$$^{(a)}$Facult\'e des Sciences Ain Chock, Universit\'e Hassan II de Casablanca;$^{(b)}$Facult\'{e} des Sciences, Universit\'{e} Ibn-Tofail, K\'{e}nitra;$^{(c)}$Facult\'e des Sciences Semlalia, Universit\'e Cadi Ayyad, LPHEA-Marrakech;$^{(d)}$LPMR, Facult\'e des Sciences, Universit\'e Mohamed Premier, Oujda;$^{(e)}$Facult\'e des sciences, Universit\'e Mohammed V, Rabat;$^{(f)}$Institute of Applied Physics, Mohammed VI Polytechnic University, Ben Guerir; Morocco.\\
% ...
$^{44}$Physics Department, Southern Methodist University, Dallas TX; United States of America.\\
% ...
$^{49}$Fakult\"{a}t Physik , Technische Universit{\"a}t Dortmund, Dortmund; Germany.\\
% ...
$^{55}$II. Physikalisches Institut, Georg-August-Universit\"{a}t G\"ottingen, G\"ottingen; Germany.\\
% ...
$^{103}$CPPM, Aix-Marseille Universit\'e, CNRS/IN2P3, Marseille; France.\\
% ...
$^{117}$$^{(a)}$New York University Abu Dhabi, Abu Dhabi;$^{(b)}$United Arab Emirates University, Al Ain; United Arab Emirates.\\
% ...
$^{121}$Homer L. Dodge Department of Physics and Astronomy, University of Oklahoma, Norman OK; United States of America.\\
% ...
$^{152}$Raymond and Beverly Sackler School of Physics and Astronomy, Tel Aviv University, Tel Aviv; Israel.\\
% ...
$^{a}$ Also Affiliated with an institute covered by a cooperation agreement with CERN.\\
$^{b}$ Also at An-Najah National University, Nablus; Palestine.\\
$^{c}$ Also at Borough of Manhattan Community College, City University of New York, New York NY; United States of America.\\
$^{d}$ Also at Center for Interdisciplinary Research and Innovation (CIRI-AUTH), Thessaloniki; Greece.\\
$^{e}$ Also at Centro Studi e Ricerche Enrico Fermi; Italy.\\
$^{f}$ Also at CERN, Geneva; Switzerland.\\
% ...

\end{flushleft}


\end{document}